\documentclass[12pt,preprint]{aastex}
\usepackage{float}
\newcommand{\gtsim}{\lower.8ex\hbox{$\; \buildrel > \over \sim \;$}}
\newcommand{\lessim}{\lower.8ex\hbox{$\; \buildrel < \over \sim \;$}}
\newcommand{\etal}{{\it et al.}}
\newcommand{\epp}{$e^+-e^--p\, $}

\shorttitle{Shock Waves in \epp\ Plasmas}
\shortauthors{Amato and Arons}

\begin{document}

\title{Heating and Non-thermal Particle Acceleration in Relativistic, 
Transverse Magnetosonic Shock Waves in Proton-Electron-Positron Plasmas}

\author{E. Amato\altaffilmark{1} and J. Arons\altaffilmark{2}}
\affil{INAF/Osservatorio Astrofisico di Arcetri\\
Largo E. Fermi, 5 - 50125 Firenze (Italy)}
\affil{Department of Astronomy, University of California, 601 Campbell hall, Berkeley, CA 94720}

\begin{abstract}
We report the results of 1D particle-in-cell simulations of ultrarelativistic shock waves in proton-electron-positron plasmas. We consider magnetized shock waves, in which the upstream medium carries a large scale magnetic field, directed transverse to the flow. Relativistic cyclotron instability of each species as the incoming particles encounter the increasing magnetic field within the shock front provides the basic plasma heating mechanism. The most significant new results come from simulations with mass ratio $m_p/m_\pm = 100$. We show that if the protons provide a sufficiently large fraction of the upstream flow energy density (including particle kinetic energy and Poynting flux), a substantial fraction of the shock heating goes into the formation of suprathermal power-law spectra of $e^--e^+$. Cyclotron absorption by the pairs of the high harmonic ion cyclotron waves, emitted by the protons, provides the non-thermal acceleration mechanism. As the proton fraction increases, the non-thermal efficiency increases and the $e^--e^+$ power-law spectra harden.

When the proton fraction is small (pair plasma almost charge symmetric), the $e^-$ and $e^+$ have approximately equal amounts of non-thermal heating. At the lower range of our simulations with mass ratio 100, when the ions contribute 56\% of the upstream flow energy flux, the pairs' non-thermal acceleration efficiency by energy is about 1\%, increasing to 5\% as the ions' fraction of upstream flow energy increases to 72\%. When the fraction of upstream flow energy in the ions rises to 84\%, the efficiency of non-thermal acceleration of the pairs reaches 30\%, with the $e^+$ receiving most of the non-thermal power.

We suggest that the varying power law spectra observed in synchrotron sources that may be powered by magnetized winds and jets might reflect the correlation of the proton to pair content enforced by the underlying electrodynamics of these sources' outflows, and that the observed correlation between the X-ray spectra of rotation powered pulsars with the X-ray spectra of their nebulae might reflect the same correlation.
\end{abstract}


\section{Introduction \label{sec:intro}}

Collisionless shock waves in relativistic astrophysical flows have been implicated as the energizing mechanism for emission in a variety of non-thermal sources - termination shocks of pulsar winds 
(\cite{slane05,kaspi04}),
hot-spot shocks terminating jets from radio galaxies (\cite{heavens87, wilson00, balcus05}) and internal shocks in AGN and microquasar jets (\cite{marscher85, turler00, turler04}), internal and external shocks in GRBs (\cite{piran04}) and outflows from other high energy transients (e.g., \cite{gaensler05}).  While shock jump conditions applied to models
of magnetohydrodynamic flow set the constraints on what fraction of flow energy goes into heat of all sorts,  the partition of that energy between ions and electrons (and positrons, when these
are present), and the distributions of particle momenta (thermal or non-thermal), are kinetic questions 
not addressed by the macroscopic conservation laws. The kinetic questions are those of collisionless plasma physics - two body encounters cannot generate the entropy implied by the jump conditions on the observed scales of transition from non-radiative flow
(upstream, in the shock excitation model)
to the observed regions of synchrotron emission (downstream, in the shock model), nor can collisional processes
lead to non-thermal particle spectra.

A fundamental approach in which the flow and kinetic structure of relativistic shocks are resolved on all scales offers the opportunity to evaluate
the nature of the electromagnetic turbulence generated, as a function of the upstream parameters of the flow, simultaneously with a determination of the heated (possibly non-thermally 
accelerated) particle spectra both up and downstream. For this purpose, fully kinetic particle-in-cell (PIC) simulations (\cite{birdsall91,dawson95}) provide a 
powerful investigatory tool, since these model the plasma from first principles, with the only approximations being those that go into the method - primarily the ``cloud-in-cell'' algorithm, which in effect softens the interactions to the grid scale.  

PIC simulations of relativistic transverse magnetosonic shocks in a symmetric pair plasma were first reported by \cite{langdon88} and by \cite{gallant92}, using spatially 1D models. 
Magnetized shocks are characterized by  coherent gyration (\cite{alsop88}) and cyclotron instability (\cite{hoshino91}) of the plasma species as 
they encounter the shock front - magnetic reflection mediates the density and velocity
transition, while the instabilty of the induced gyration mediates the entropy generation. Simulations of the Weibel instability when $B_0 =0$ with applications to unmagnetized relativistic shocks have been reported 
recently by a number of authors (\cite{silva03,fred04,hed04,jaro05,hed05}). 
Spitkovsky and 
Arons (in preparation) have found comparable results, and show how the shock 
structures in pair plasmas depend on the magnetic field strength and 
orientation with respect to the flow.  All these results indicate that 
non-thermal particle acceleration is either weak or absent in shocks in pair 
plasmas.

\cite{hoshino92} found that when the upstream flow includes heavy ions which carry a significant fraction of the upstream flow energy, non-thermal acceleration {\it does} occur, as a consequence of the absorption of ion waves emitted at harmonics of the ion cyclotron frequency reabsorbed by the pairs, even though the shock structure is one dimensional and no cross field diffusion occurred. The interaction is resonant, in contrast to the non-resonant scattering underlying Diffusive Shock Acceleration (DSA). Computational limitations required the use of a low rest mass ratio, $m_i /m_\pm = 10, 20$. 

Those inhomogeneous shock calculations showed no electron acceleration. While most of the characteristics of the instability scale with just the energy density ratio between the protons and the pairs, the efficacy of the waves as accelerators of electrons and positrons also depends on the polarization properties of the proton cyclotron waves.
The lack of electron acceleration at low mass ratio stems from the polarization of the waves excited. In a quasi-neutral pair plasma the waves are linearly polarized. If the heavy ions are a small fraction by number, the waves excited by the ion maser remain almost linearly polarized. Since linearly polarized waves are an equal mixture of left and right circularly polarized modes, equal electron and positron acceleration might have been expected. However, the simulations also revealed that the mechanism is a significant accelerator only if the energy density in the ions $U_{1i} = m_i N_{i1} ( \gamma_1 -1)  c^2$, where $N_{i1}$ is the upstream ion density 
and $\gamma_1$ is the upstream flow Lorentz factor, is comparable to the upstream energy density in the pairs, $U_{1\pm} = m_\pm ( \gamma_1 -1) c^2 (N_{1+} + N_{1-})$. Then 
$N_{1i}/(N_{1+} + N_{1-}) = (m_\pm /m_i) (U_{1i}/U_{1\pm}) \sim m_\pm /m_i$. For the small mass ratios
employed, the dispersion relations for small amplitude waves show that at comparable energy densities
in the species, the waves are still strongly left handed, with consequent preference for positron acceleration.

The earlier work also entirely neglected upstream thermal spread in the momenta of the species. The acceleration mechanism relies upon cyclotron absorption of waves in the pairs at high harmonics of the ion cyclotron 
frequency. Since the shock in the pairs alone heats the pairs to a temperature 
$T_{2\pm} \approx m_\pm c^2 \gamma_1$ (if $\sigma_1 \equiv B_1^2/4\pi \rho_1 \gamma_1 c^2 \ll 1 $, with $B_1$
the upstream magnetic field strength and $\rho_1$ the upstream rest mass density, both measured in the downstream frame), 
efficient non-thermal acceleration of the pairs requires cyclotron absorption of ion waves at frequency 
$\omega \approx \Omega_{c\pm} = eB /m_\pm c \gamma_1 = (m_i /m_\pm ) eB /m_i c \gamma_1
= (m_i /m_\pm ) \Omega_{ci}$; that is, the relativistic cyclotron instability of the ion ring in the shock front's leading edge must generate significant power at harmonic numbers $ n \sim m_i /m_\pm \gg 1$. 
When the upstream plasma is completely cold, this is not a problem - the growth rate of the instability in 
a homogeneous medium scales as $\Omega_{ci} n^{-1/9}$ (\cite{hoshino91}).
We shall see that if there is thermal dispersion in the upstream ion momenta $\Delta p $, 
growth is inhibited, with no substantial wave power at harmonic numbers above 
$m_i c \gamma_1 /\Delta p$.

In this paper we reinvestigate the one dimensional shock structure of a transverse 
magnetosonic shock, with particular attention to the effect of more realistic mass 
ratios and the incorporation of thermal dispersion in the upstream medium. 

In \S \ref{sec:shocks} we undertake the study of the structure and acceleration 
properties of transverse relativistic shocks.  \S \ref{sec:pairshocks}  
contains a study of shocks in pure pair plasma. Consistent with previous results,
we find no signs of non-thermal particle acceleration. We introduce a proton 
component in \S \ref{sec:dens} and find, for the first time in the literature, 
evidence of non-thermal acceleration not only of positrons but also of electrons. 
In \S \ref{sec:high-mass} we show that at 
high mass ratio the waves become close to linearly polarized when the ion 
and pair energy densities are comparable. This leads to comparable heating and 
acceleration rates for positrons and electrons. In \S \ref{sec:temp} we include 
in the shock simulations a thermal spread of the ions' upstream distribution and 
study how this affects the results. Finally, we present a summary of our results 
and our conclusions in \S \ref{sec:discuss}. Aspects of the linear stability of
relativistic protons gyrating in a relativstically hot pair plasma are described
in the Appendix, \S \ref{sec:linear}.

\section{Structure and Non-thermal Particle Acceleration Properties of Relativistic,  Transverse Magnetosonic  Shocks \label{sec:shocks}}

We study relativistic transverse magnetosonic shock
waves. Our analysis was carried out through numerical simulations performed
with XOOPIC (\cite{xoopic}). The simulations were 1D and were performed 
in the geometry shown in Fig.~\ref{fig:shockgeom} below. 

When the simulation starts, the simulation box is permeated with a 
static uniform magnetic field directed along the $z$-axis and a
uniform electric field along the $y$-axis. New plasma is constantly
injected in the simulation box from the left boundary. The plasma 
moves along the $x$-axis with velocity
${\bf v}_{x1}$ (four-velocity ${\bf u}_{x1}$). The drift motion is initialized and maintained
in the upstream medium by ${\bf E}_y$, which
is set equal to 
${\bf E}=-{\bf v}_{x1}/c \times {\bf B}$.

The right boundary of the simulation box is a conductor which behaves
as a perfectly reflecting wall, both for particles and electromagnetic 
radiation. When the plasma first impacts on the wall, a reverse shock 
propagates towards the left. What we are interested in is the behaviour 
of the fluid at, and right after, the crossing of this reverse shock.  

\begin{figure}[H]
\resizebox{15cm}{!}{
\includegraphics[angle=-90]{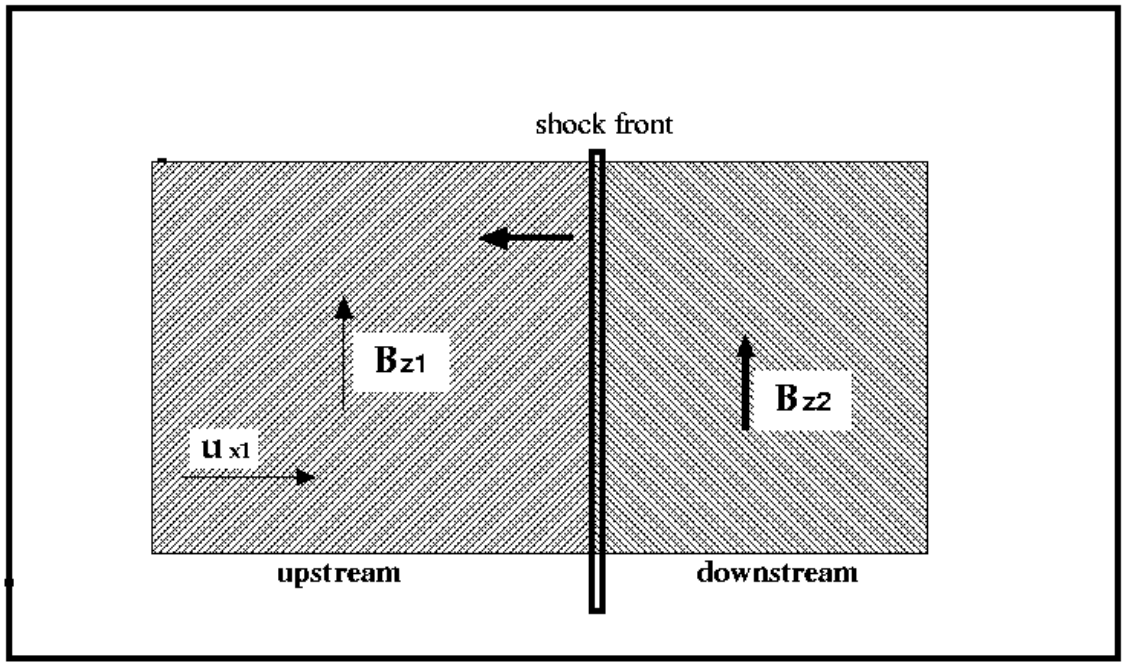}}
\caption{The simulation geometry.}
\label{fig:shockgeom}
\end{figure}

The upstream and downstream conditions are related by the
ideal MHD jump conditions for a relativistic, transverse, magnetosonic shock.
These were derived
by \cite{kc1}, as expressed in the shock frame, and by 
\cite{gallant92} as expressed in the downstream frame. 
Quantities
expressed in this latter frame are more convenient for purposes of
direct comparison of the results of our numerical simulations
with the theoretical expectations.
Since the wall on the right of the simulation
box is at rest, the $x$-component of the fluid velocity must vanish
downstream of the shock. The downstream frame is 
therefore coincident with the computational observer's frame and it is the frame in
which all quantities that will enter the following discussion are
measured. In this frame, the shock moves toward the left, 
with velocity $-c \beta_{shock}$.
We assume the fluid upstream to be cold,
$p_1 \ll n_1 m c^2$, where $p_1$ and $n_1$ are the proper pressure and density
upstream, and to move at ultrarelativistic speed, with Lorentz factor
$\gamma_1 \gg 1$ and 4-velocity $c u_1\approx c \gamma_1$. For the fluid downstream,
we assume the ideal ultrarelativistic equation of state to hold, with proper
enthalpy $w_2 \approx \Gamma p_2/(\Gamma-1)$. The adiabatic index
$\Gamma$ would be equal to $4/3$ for an isotropic relativistic fluid.
But in our analysis the particles' momenta are restricted to the plane
perpendicular to ${\bf B}$ therefore the appropriate value of $\Gamma$
is $3/2$, describing a 2-dimensional ultrarelativistic fluid.

After expressing the densities
in terms of the simulation frame values
$N=n \gamma$, the jump conditions reduce to: 
\begin{eqnarray}
{N_2 \over N_1}&=&{1+\beta_{\rm shock} \over \beta_{\rm shock}}\ ,
\label{eq:num3}\\
& & \nonumber \\
\left(1+\beta_{\rm shock} \right) N_1 m \gamma_1 c^2 (1+\sigma)&=&
\beta_{\rm shock} \left({p_2 \over \Gamma-1}+{B_2^2 \over 8 \pi}  \right)\ , 
\label{eq:en3}\\
& & \nonumber \\
\left(1+\beta_{\rm shock} \right) N_1 m \gamma_1 c^2 (1+\sigma)&=& 
p_2+{B_2^2 \over 8 \pi}\ , \label{eq:mom3}\\
& & \nonumber \\
{B_2 \over B_1}&=&{1+\beta_{\rm shock} \over \beta_{\rm shock}}\ ,
\label{eq:etan3}
\end{eqnarray}
where we have used the definition of the magnetization parameter
$$\sigma={B_1^2 \over 4 \pi N_1 m c^2 \gamma_1}.$$ 

Using Eq.~\ref{eq:mom3} and Eq.~\ref{eq:etan3} to
eliminate $p_2$ and $B_2$ from Eq.~\ref{eq:en3}, one finds
\begin{equation}
\label{eq:shocksol}
\beta_{\rm shock}= {1 \over 2 (1+1/\sigma)}
\left\{ \left({\Gamma \over 2 }+
{\Gamma-1 \over \sigma}\right)+ \left[
\left({\Gamma \over 2}+{\Gamma-1 \over \sigma}\right)^2
+4 \left(1-{\Gamma \over 2}\right)
\left(1+{1 \over \sigma} \right) \right]^{1/2} \right\}\ .
\end{equation} 
The jump in all quantities at the crossing of the shock front is then
determined as a function of $\sigma$ and $\Gamma$. The density and
magnetic field compression are determined through Eqs.~\ref{eq:num3} 
and \ref{eq:etan3}, while the ratio between the
thermal pressure downstream and the ram pressure upstream can be
similarly expressed in terms of $\sigma$ and $\beta_{\rm shock}$
after manipulation of Eq.~\ref{eq:en3}. One obtains:
\begin{equation}
\label{eq:temp3}
{k T_2 \over m \gamma_1 c^2}=\beta_{\rm shock} 
\left[1 + \sigma \left(1-{1+\beta_{\rm shock} \over 2 \beta_{\rm shock}} \right) \right]
\end{equation}

The low and high $\sigma$ limits ($\sigma \ll 1$ and $\sigma \gg 1$, 
respectively) of Eqs.~\ref{eq:num3}, \ref{eq:etan3}, \ref{eq:shocksol}
and \ref{eq:temp3}
are discussed extensively by \cite{gallant92}, and the main
result is that with increasing magnetization the fraction of 
upstream energy that goes into an increase of the magnetic pressure
downstream becomes larger, at the expense of the particle thermal
energy. In the following, we shall deal with the case of a plasma
with $\sigma \approx 1$.

\subsection{Numerical simulations of relativistic transverse shocks in 
a pure pair plasma \label{sec:pairshocks}}

The results we show and discuss in the following refer to a simulation
performed on a plasma with $\sigma_{-}=\sigma_{+}=2$, corresponding to
$\sigma_{tot} = 1$, and Lorentz 
factor upstream $\gamma_1=40$. The simulation grid was made of 1024 cells, 
each of size $\Delta x=r_L/10$, with $r_L$ being the particle 
Larmor radius based on the upstream $B$ field and flow Lorentz factor. 
The initial number of particles per cell was 32 for
each species. The results are basically unchanged when $\Delta x$ is
varied from $r_L/10$ to $r_L/5$, the number of cells between 512 and
2048, and the number of particles of each species per cell from 8 to
64. The time-step is $\Delta t =0.9 \Delta x/c$ in all cases, in order
to satisfy the Courant condition. 
\begin{figure}[H]
\resizebox{\hsize}{!}{
\includegraphics{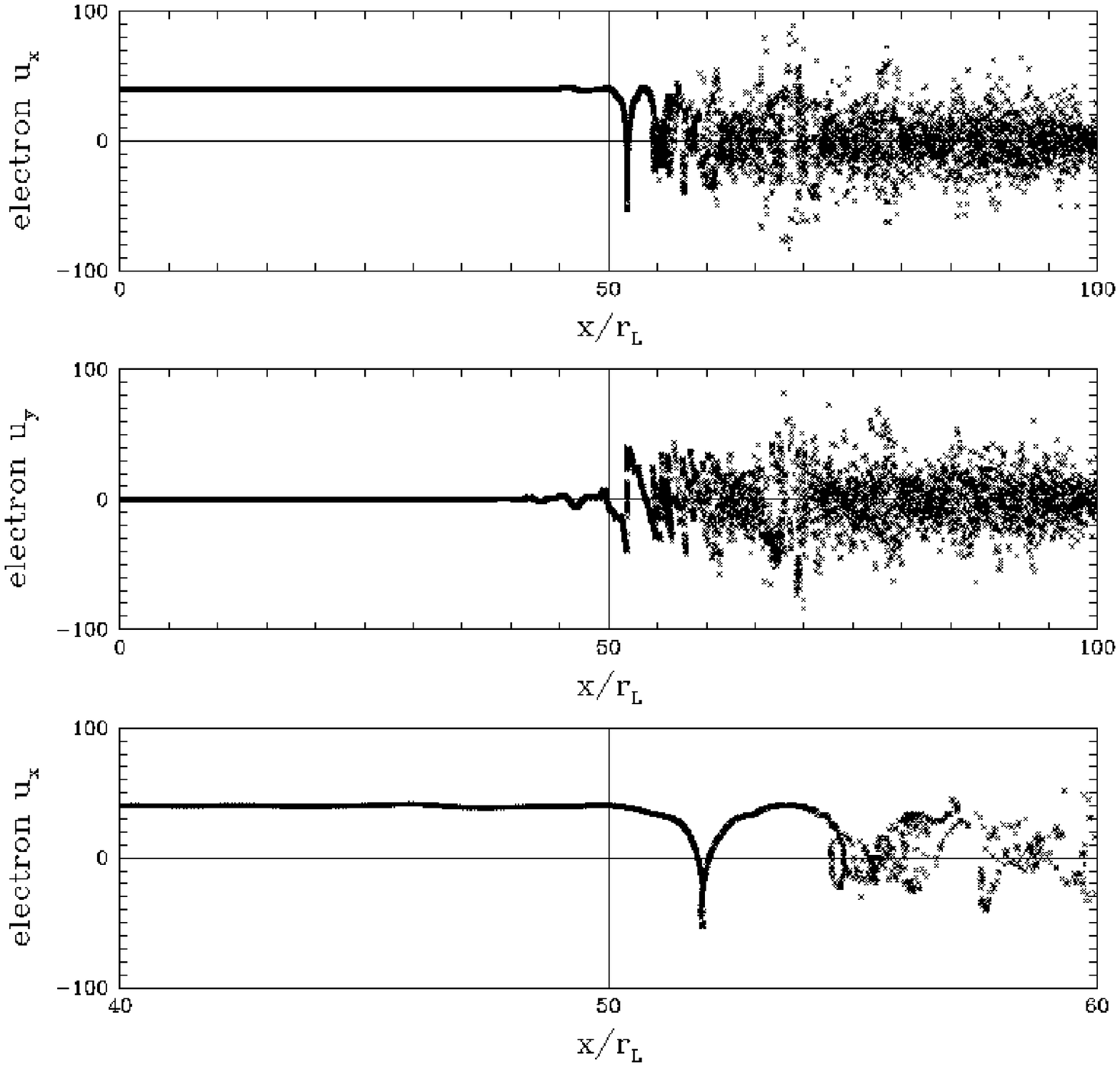}
\includegraphics{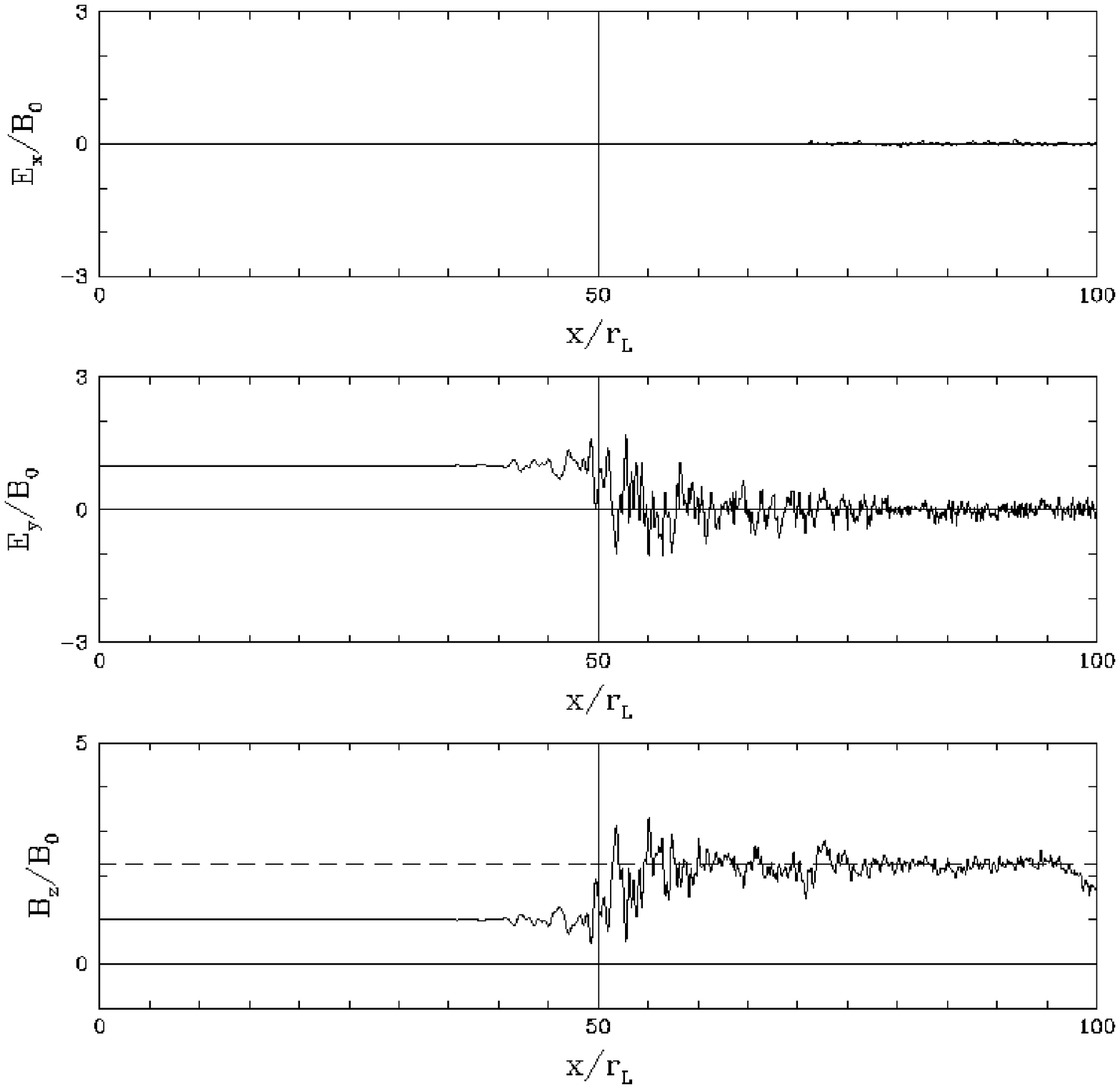}}
\caption{{\bf Plots on the left:}
The electron four-velocity as a function of position
in the box. The x-coordinate is in units of $r_L$, the pair Larmor radius
upstream. The shock front is located at $x_s/r_L \approx 50$.
Top panel: x-component of the four velocity. Middle panel:
y-component of the four-velocity. Bottom panel: a zoom of the top panel
onto the region around the shock front. {\bf Plots on the right:}
The electromagnetic field as a function of 
position in the box. The x-coordinate is again in units of the pair
Larmor radius upstream. Top panel: longitudinal component of the
electric field. Middle panel: transverse component of the electric 
field. Bottom panel: the magnetic field.}
\label{fig:pairvel}
\end{figure}
As it was the case for the simulations performed by \cite{gallant92},
we find in our simulation the presence of an electromagnetic precursor
wave, propagating ahead of the shock, into the upstream medium, at 
essentially the speed of light.
The results in the following figures refer to a time when the reverse
shock has propagated far enough from the right wall of the simulation,
so that edge effects (spurious oscillations due to the vicinity of the
conducting boundary) do not affect the dynamics around the shock front,
while the precursor has not reached the left boundary
of the box yet.

\begin{figure}[H]
\resizebox{\hsize}{!}{
\includegraphics{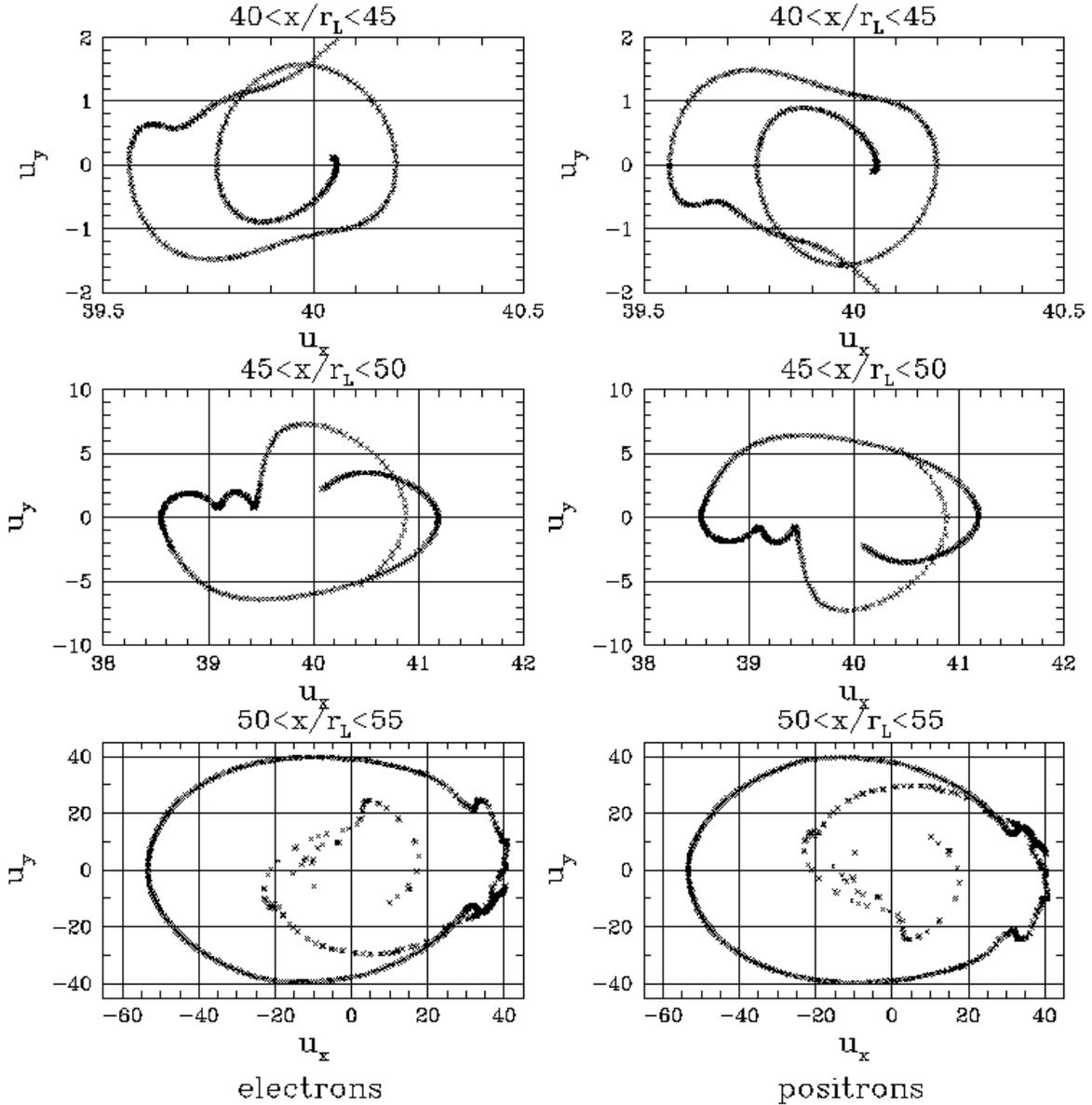}}
\caption{The electron (panels on the left) and 
positron (panels on the right) distribution in $u_x-u_y$ space
in different regions around the shock front, as specified on top of
each panel.}
\label{fig:pairring}
\end{figure}

In Fig.~\ref{fig:pairvel} we show, on the left, the ($u_x-x$) and ($u_y-x$) 
sections of the electrons' phasespace. The behavior of the positrons is
completely analogous. The main thing to notice
is the loop that develops at the crossing of the shock in the ($u_x-x$)
plane, especially evident in the bottom panel. This is the initial 
reflection loop described by the soliton solution of \cite{alsop88}. 

The particles' four-velocities in the reflection region form 
distorted cold rings 
in momentum space, as shown in Fig.~\ref{fig:pairring}, so that
one can expect cyclotron instability to develop, as expected from
the uniform medium instability theory outlined in the Appendix
with the ions omitted (see also \cite{hoshino91}).

In Fig.~\ref{fig:pairvel} we also plot the different electromagnetic
field components as a function of position. The longitudinal electric
field always stays around zero with negligibly small oscillations. 
This is just what one expects due to the charge and mass symmetry,
together with overall neutrality, leading to decoupling of the 
transverse from longitudinal extraordinary mode.

The transverse component of the electric field vanishes on average
behind the shock, as predicted by ideal MHD, but of course there
are oscillations due to the development of cyclotron instability
in the pairs. The same behavior is seen in the magnetic field.
The latter, after the first two main peaks, corresponding to the turning
points of the first reflection loop, oscillates around the value 
computed from the jump conditions 
(Eqs.~\ref{eq:etan3} and \ref{eq:shocksol}) for a plasma 
with $\sigma=1$ and $\Gamma=3/2$ (dashed curve in the plot). 

\begin{figure}[H]
\resizebox{\hsize}{!}{
\includegraphics{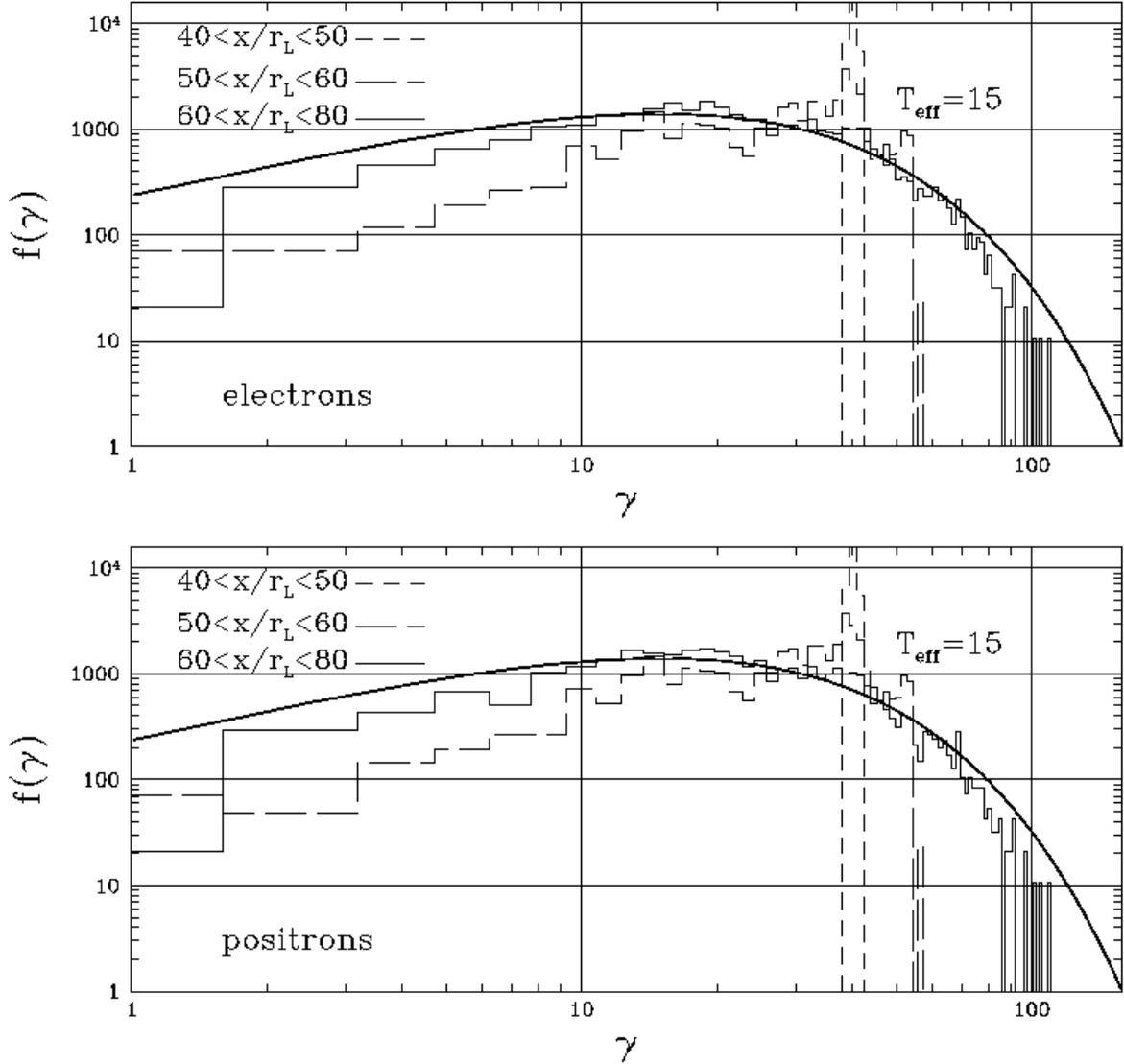}}
\caption{The evolution of the particle distribution after
the crossing of the shock. The different line-types refer to particles 
extracted from different slices of the simulation domain as specified in the
figure: short-dashed for particles at
positions $40<x/r_L<50$, long-dashed for positions $50<x/r_L<60$ and
solid for positions $60<x/r_L<80$. The top panel refers to electrons 
and the bottom
one to positrons. In both cases $f(\gamma)$ is normalized to the particle
number denisty upstream. The thick solid curve represents a 2-D relativistic 
maxwellian of temperature $T_{\rm eff}$ (see text).}
\label{fig:pairdistr}
\end{figure}

Finally, in Fig.~\ref{fig:pairdistr} we show the evolution of the
pair distribution function at the crossing of the shock. It is 
apparent that no signs of non-thermal acceleration 
are observed. The distribution function progressively evolves towards
a maxwellian for both species. The temperature of the maxwellian is
in perfect agreement with the prediction of ideal MHD (Eq.~\ref{eq:temp3}): 
$T_{eff}=kT_{2}/ m_\pm c^2 \approx 0.4\ \gamma_1$.

\subsection{Numerical Simulations of relativistic transverse shocks 
in $e^--e^+-p$ plasmas: acceleration of electrons \label{sec:dens}}

We show that with protons present in the upstream flow, the downstream
$e^\pm$ exhibit non-thermal heating, a.k.a. shock acceleration.
We first 
tackle the problem of how the particle acceleration process is
affected by the waves' polarization. 

In the following we discuss the results of two simulations in which a 
plasma with $\sigma_{-}=2$ and $\sigma_i=0.25$ is considered. 
The latter condition is realized
in two different ways, first adopting a mass and number ratio between ions
and electrons equal to 20 and 0.4 respectively (case A), and then using
$m_i/m_\pm=40$ and $N_i/N_{-}=0.2$ (case B).

The simulation geometry is that in Fig.~\ref{fig:shockgeom}.
In case A the simulation grid contains 2048 cells, each of size $r_{L1\pm}/10$,
where $r_{L1\pm} = m_\pm c^2 \gamma_1 /e B_1$, while in case B the cell 
size is the same but the number of cells is doubled.
In both cases the box contains more than 10 ion Larmor orbits and the
number of particles per cell is 16. The initial
plasma Lorentz factor is $\gamma_1=40$, but no dependence of the 
results on 
this parameter is expected as long as the condition $\gamma_1 \gg 1$ is
satisfied. As to the time-step, finally, this is taken to be
$\Delta t=0.9 \Delta x/c$, so that the Courant condition is ensured.

\begin{figure}
\resizebox{\hsize}{!}{
\includegraphics{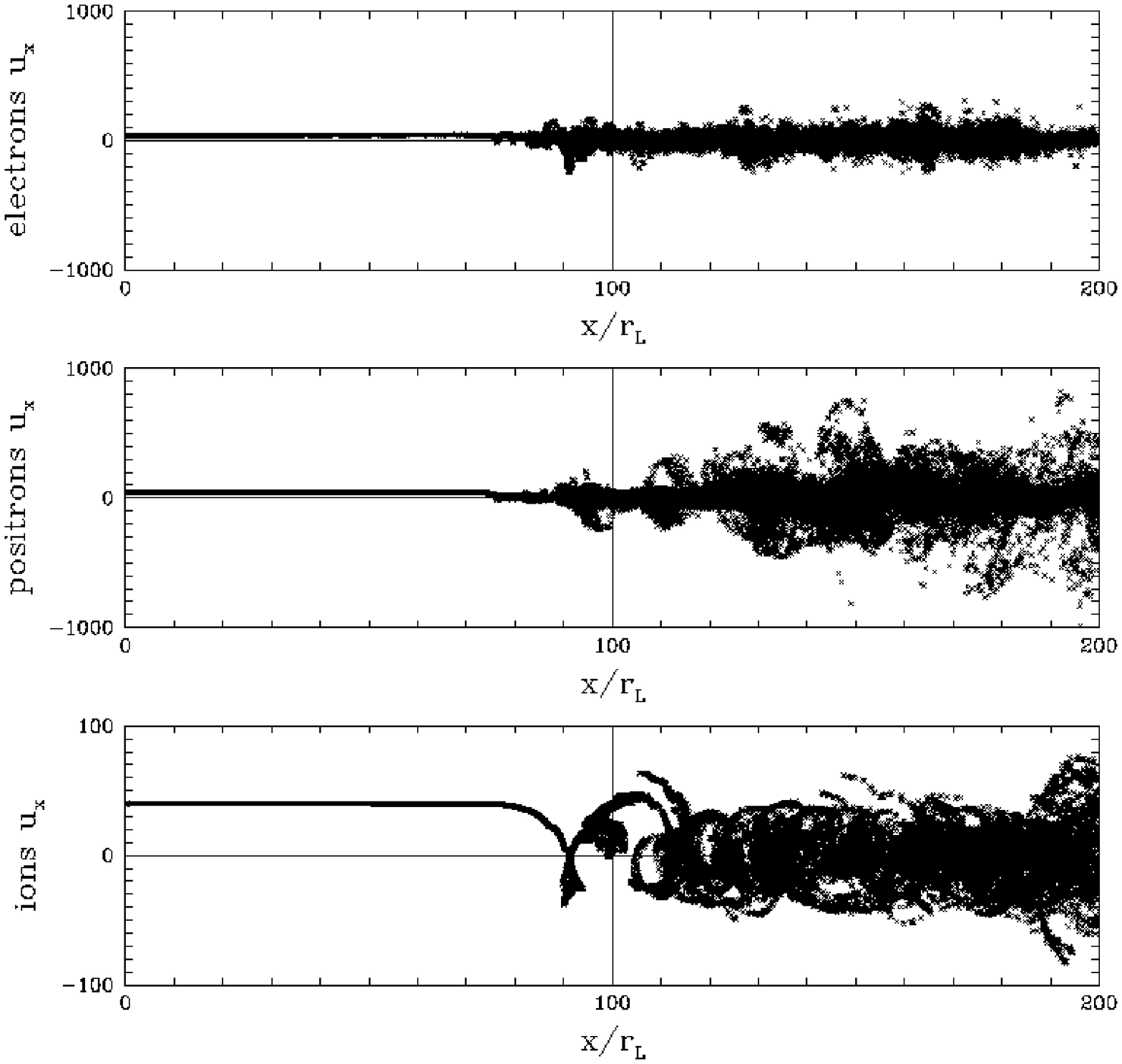}
\includegraphics{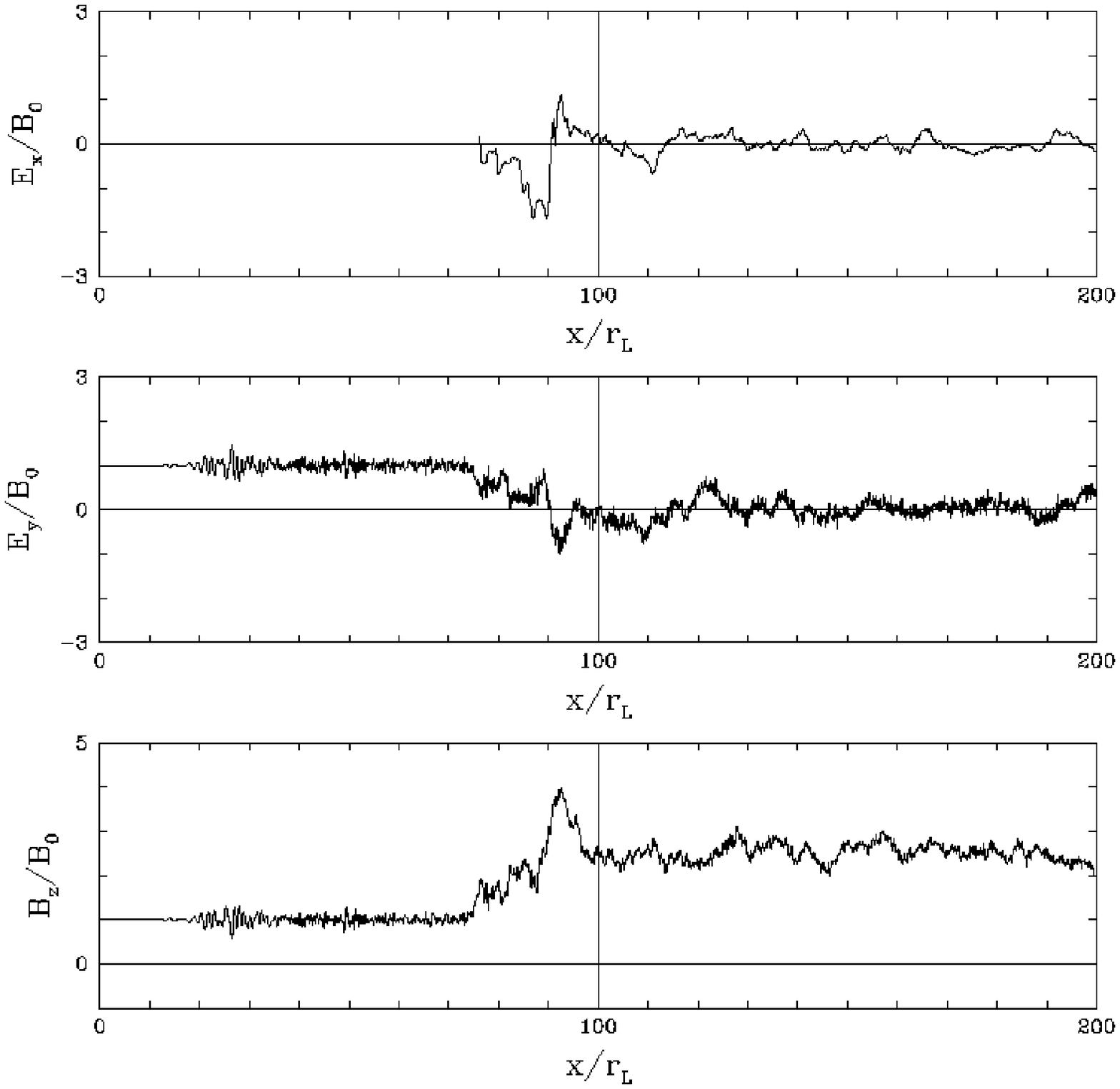}}
\caption{{\bf Left panel:} The ($u_x-x$) section of the 
phase space of the different species for case A: electrons in the top panel, 
positrons in the middle panel and ions in the bottom panel. 
{\bf Right panel:} The electromagnetic field components in the simulation
labeled A: the fields are normalized to the upstream magnetic field intensity.
In all plots the $x$-coordinate is in units of the pair gyration radius
upstream.}
\label{fig:epp20}
\end{figure}

\begin{figure}
\resizebox{\hsize}{!}{
\includegraphics{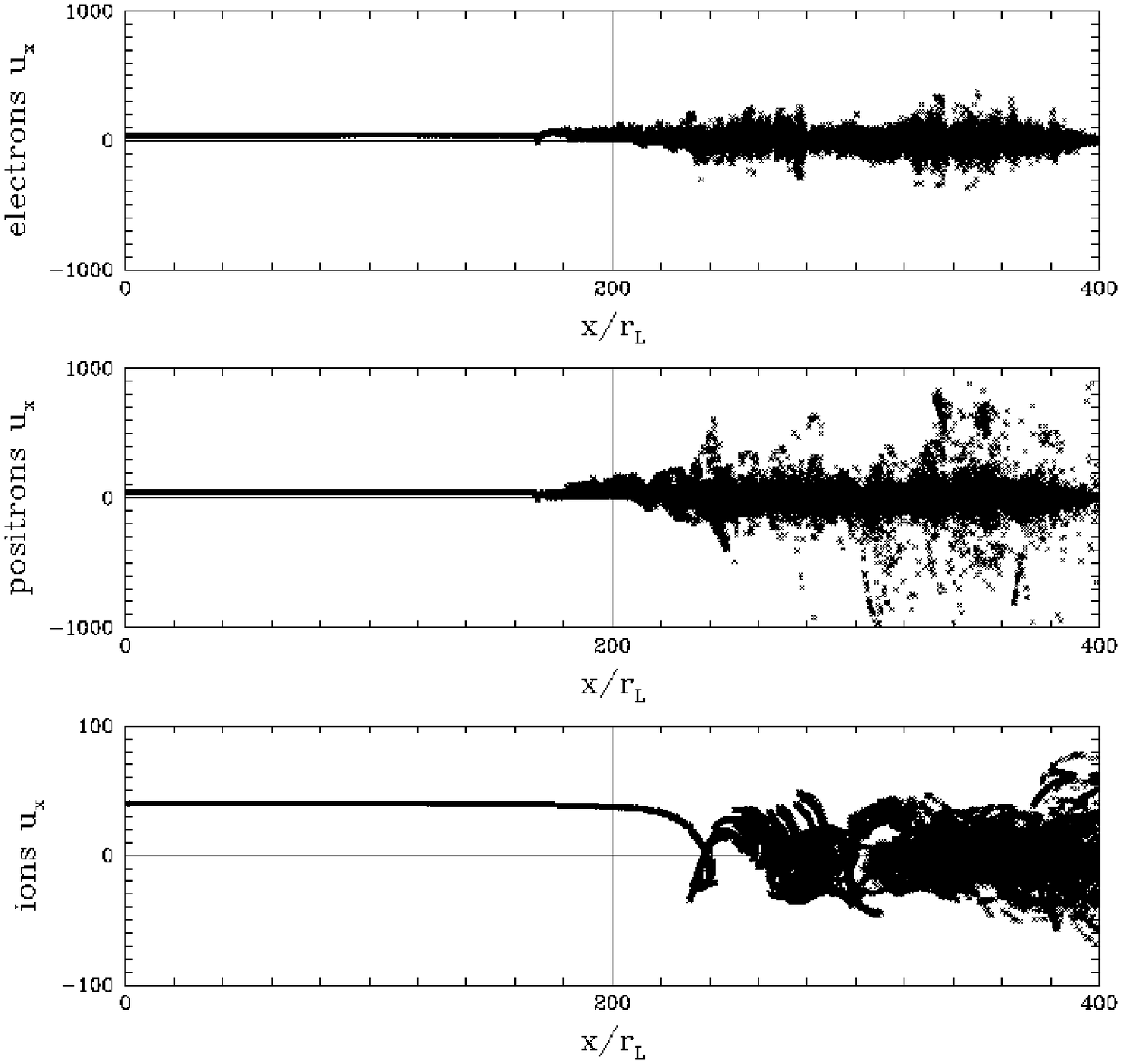}
\includegraphics{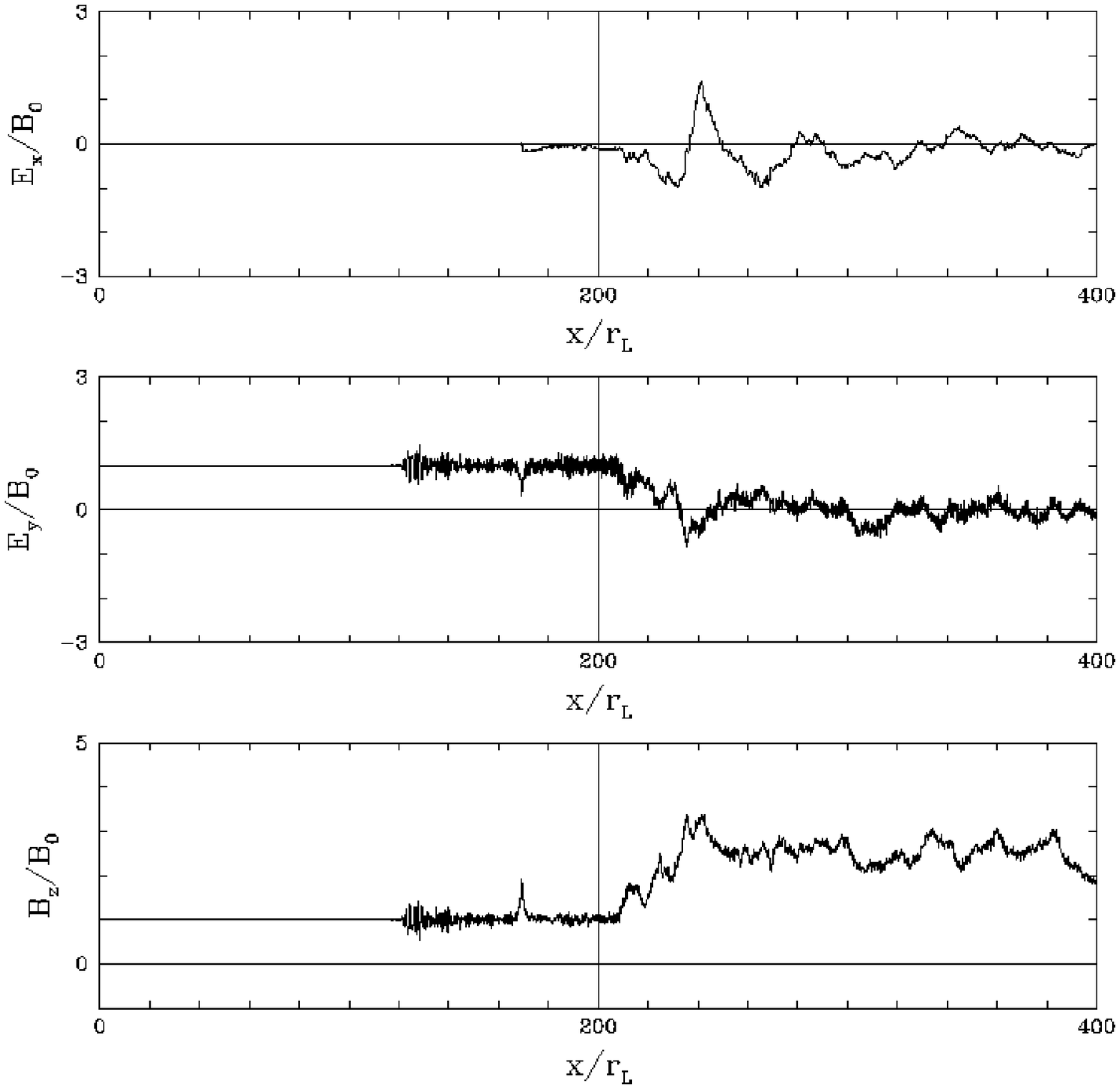}}
\caption{Same as Fig.~\ref{fig:epp20} but for case B.}
\label{fig:epp40}
\end{figure}

In Figs.~\ref{fig:epp20}-\ref{fig:epp40} we show, on the left, 
the section ($u_x-x$) of the phase-space of the different species 
in the two different cases. The reverse shock is at around $x_s\approx 100 r_L$
in Fig.~\ref{fig:epp20} and $x_s\approx 200 r_L$ in Fig.~\ref{fig:epp40}.

Comparing the left panels of the two figures, the first 
ion reflection loop is well defined 
in both cases, and the overall velocity profiles look very 
similar. At a closer look, however, one notices that in case B 
(Fig.~\ref{fig:epp40}) the spread in the
electron four-velocity post-shock is larger than in case A,
while the opposite happens for the positron velocities. This is a 
first signature of what we will find to be the most important difference 
between the two simulations.

Before discussing this point, however, let us briefly comment on the
plots showing the electromagnetic field components in the two
cases. First of all, in both cases, the presence of the electromagnetic 
precursor propagating ahead of the shock in the upstream medium is well 
evident. This wave 
is a purely transverse electromagnetic wave, as one can readily deduce
from the absence of fluctuations in the longitudinal component of the
electric field upstream of the shock. It is the same precursor as appears in 
magnetized shocks in pure pair plasmas (see previous section and also
\cite{gallant92}), and is emitted 
by the quasi-coherent rings of $e^\pm$ reflected at the leading edge of 
the whole shock structure.

\begin{figure}
\resizebox{\hsize}{!}{
\includegraphics{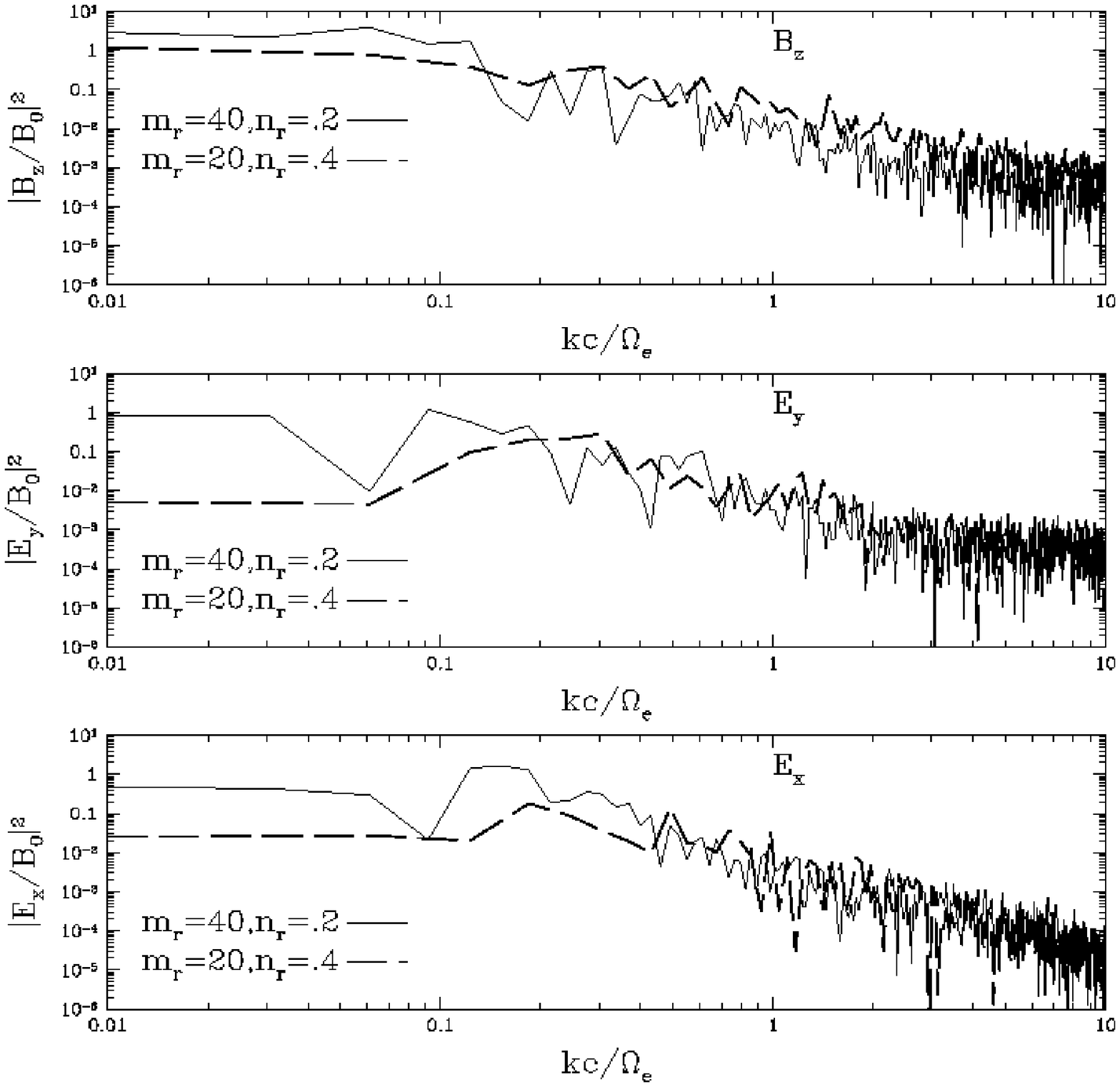}}
\caption{The downstream power spectra of the different 
electromagnetic field components, for case A (dashed curves) and case B
(solid curves). The power spectrum is normalized to the value of the
magnetic field upstream, while the wavenumbers are in units of 
$\Omega_{ce}/c$, with $\Omega_{ce}$ being the pairs' Larmor frequency 
upstream.}
\label{fig:compshock}
\end{figure}

As to the fields downstream, now the oscillations observed in the 
longitudinal component of the
electric field are larger than in Fig.~\ref{fig:pairvel}: 
due to the fact that the charge 
and mass symmetry of the pure pair plasma is broken by the presence 
of the ions, the transverse mode is partly longitudinal. 
The magnetic field oscillates around the enhanced post-shock value 
predicted by ideal MHD. The properties of the fields downstream 
are similar in the two cases we are now 
considering as it is clear from Fig.~\ref{fig:compshock},
where we compare the downstream power spectra extracted from
the two simulations.

In both cases one finds that the power as a
function of frequency can be approximated around the pair cyclotron
frequency as a power law with index -2.
\begin{figure}[H]
\resizebox{\hsize}{!}{
\includegraphics{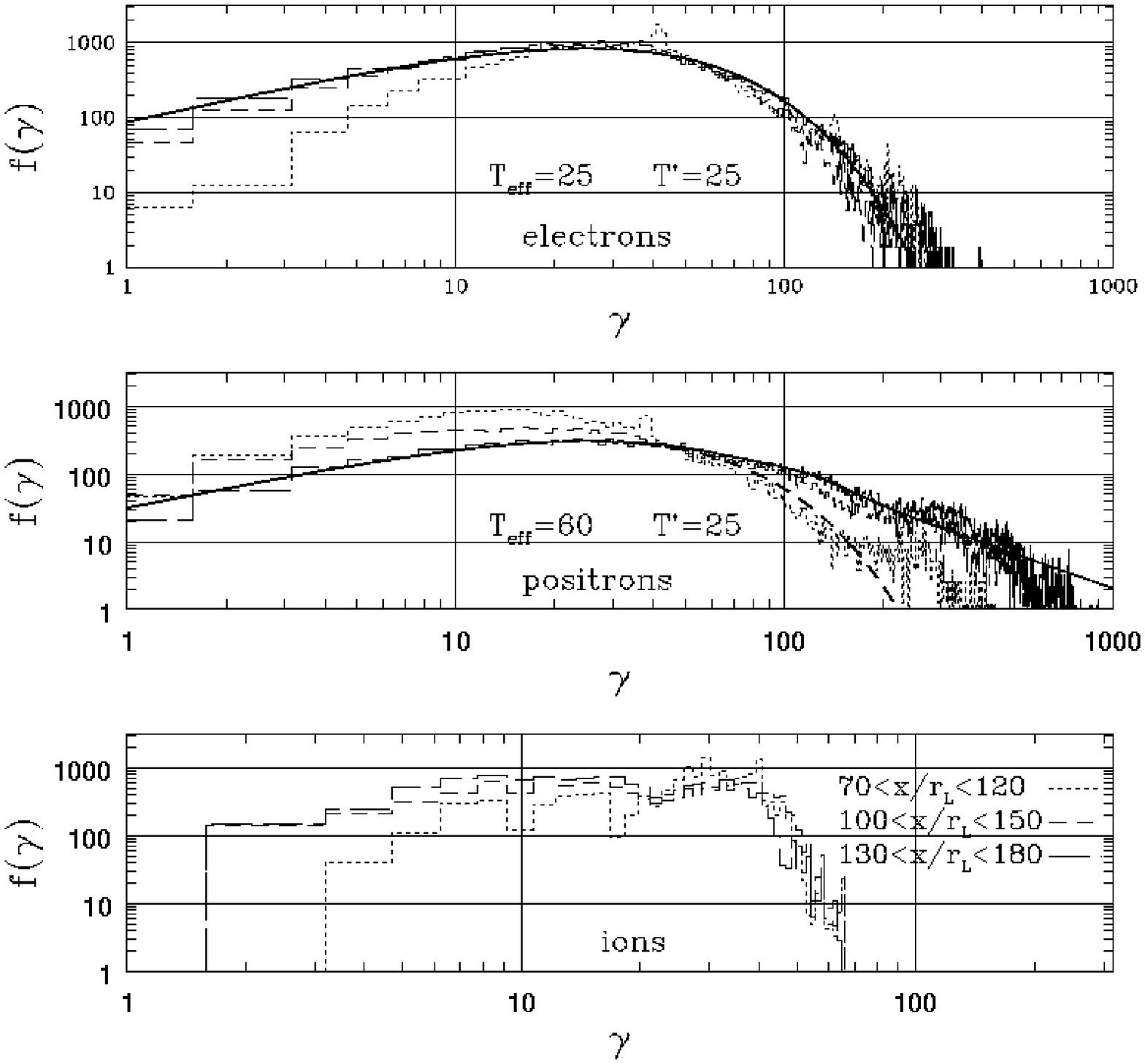}
\includegraphics{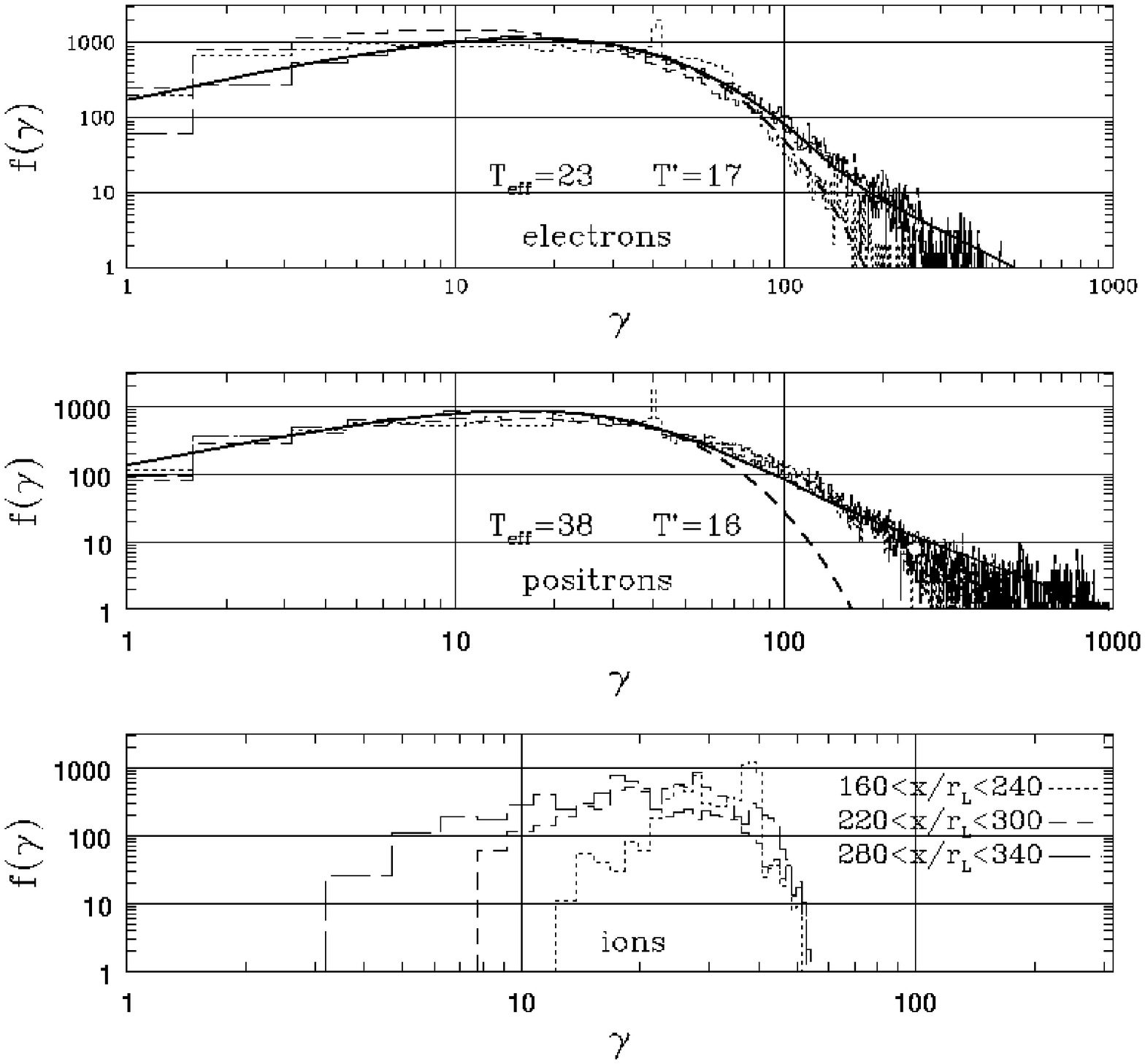}}
\caption{The postshock evolution of the distribution function
of the different species, each normalized to the corresponding 
upstream density. 
The left panels refer to case A and the right panels to 
case B. The different line-types refer to different ranges of particle 
positions, as speciefied in the lower panel of each column.
For electrons and positrons (upper and middle panels of each column) two
fits of the distribution functions are presented together with the data.
The thick dashed curve is a 2-D relativistic maxwellian with temperature $T'$,
while the thick solid curve is a maxwellian plus power-law fit (see text).}
\label{fig:allhists}
\end{figure}

Finally we discuss the postshock evolution of the particle distribution
functions. In Fig.~\ref{fig:allhists} we plot for both cases (case A 
on the left and case B on the right) the distribution function
of the different species within different regions behind the shock.

We observe, for all species, and in all cases, a progressive broadening 
of the distribution function with increasing distance from the shock front. 
However the details of the process are different between the different 
species and between the two simulations.

In case A, the energy release of the ions is quicker than in case B. 
One also sees, in the plots on the left, 
that a preferential heating of
positrons occurs, as shown by the effective temperatures of the 
different species (value of $T_{\rm eff}$ reported in the plots). 
Moreover, positrons are not only heated but
also non-thermally accelerated, since the energy excess is found in 
the form of a well developed high energy suprathermal tail.

In case B, one still observes hotter positrons
than electrons, but the difference between the effective temperatures
of the two species is smaller. The final value of $T_{\rm eff}$ is,
for both species, smaller than in case A, and the suprathermal positron
tail is reduced, despite being still evident, while signs of 
electron acceleration are found for the first time.  

We fit the final distribution functions of the pairs
(thick solid curves in the top and middle panels of Fig.~\ref{fig:allhists})
with a function $N(\gamma)$ given by the superposition of
a 2-D relativistic maxwellian plus a power-law tail, with the 
latter exponentially cut-off at low energies:
\begin{equation}
\frac{N_\pm(\gamma)}{N_{1\pm}}={f \over T' (T'+1)} \gamma \exp{\left[-{\gamma-1 \over T'}\right]}+
N_t \left( {\gamma \over \gamma_m} \right)^{-\alpha} 
\left\{1- \exp{\left[-\left({\gamma-1 \over \gamma_m}\right)^r\right]} \right\}\ .
\label{eq:fitdistr}
\end{equation} 
In Eq.~\ref{eq:fitdistr}, $N_1$ is the upstream density of the 
species considered and $T' = T/m_\pm c^2$, 
$\alpha$ and $r$ are the power-law indices of the
suprathermal tail and of the cut-off, respectively, and, finally, 
$\gamma_m$ is the Lorentz factor at which the power-law tail is cut-off. 

The best fit function is plotted in each of the relevant panels of 
Fig.~\ref{fig:allhists} as a thick solid curve, while the thick dashed
curve represents a maxwellian of temperature $T' \equiv T/m_\pm c^2$. 
The curves we plot are defined by using for the parameters appearing in 
Eq.~\ref{eq:fitdistr} the values listed in Table \ref{tab:epp2040}. 
In the last row of the same table we report the acceleration efficiency 
in terms of energy, $\eta$. This is computed as the ratio between the 
energy density of 
$e^-$ and $e^+$ in the suprathermal tail of the distribution, $U_{nt\pm}$
and the total upstream flow energy density $U_{tot}$:
\begin{equation}
\eta_\pm=U_{nt\pm}/U_{tot}\ ,
\label{eq:eta}
\end{equation}  
with 
\begin{eqnarray}
U_{tot}&=&(\gamma_1-1)c^2 \left[N_{1i} m_i+m_\pm(N_{1-}+N_{1+})\right]+
\frac{B_1^2}{8 \pi}(1+\beta_1^2)\nonumber\\
&=&(\gamma_1-1)m_\pm c^2 N_{1-}\left[2 +\left(\frac{m_i}{m_\pm}-1\right)\frac{N_{1i}}{N_{1-}}+\sigma_{-}\right]\ ,
\label{eq:utot}
\end{eqnarray}
and $U_{nt\pm}$ computed as the integral of the second term in 
Eq.~\ref{eq:fitdistr} multiplied by $m_\pm \gamma c^2$:
\begin{equation}
U_{nt\pm}=N_{1\pm} m_\pm c^2 \gamma_m^2 N_t \int_1^{\gamma_{max}/\gamma_m}dx\ 
x^{1-\alpha}\ 
\left\{1-\exp\left[-\left(x-\frac{1}{\gamma_m}\right)^r\right]\right\}\ .
\label{eq:unt}
\end{equation}
In the latter equation we see that a parameter $\gamma_{max}$ appears, that 
refers to the maximum Lorentz factor up to which particles are accelerated.
The theoretical upper limit for $\gamma_{max}$ within the framework of the 
proposed acceleration mechanism is $\gamma_{max}^{th}=m_i/m_\pm \gamma_1$,
i.e. the pairs cannot be accelerated up to energies exceeding the ions' energy
upstream.
\begin{table}[H]
\begin{center}
\begin{tabular}{||c||c|c|||c|c||}
\hline
\multicolumn{5}{||c||}{Simulation parameters}\\
\hline
&\multicolumn{2}{c|||}{{\bf case A}} &\multicolumn{2}{|c|||}{{\bf case B}}\\
\hline
\hline
$m_i/m_\pm$&\multicolumn{2}{c|||}{$20$} &\multicolumn{2}{|c||}{$40$}\\
\hline
$N_i/N_{-}$&\multicolumn{2}{c|||}{$0.4$} &\multicolumn{2}{|c||}{$0.2$}\\
\hline
$U_i/U_{tot}$ &\multicolumn{2}{c|||}{$0.69$} &\multicolumn{2}{|c||}{$0.68$}\\
\hline
$U_i/U_{-}$ &\multicolumn{2}{c|||}{$8$} &\multicolumn{2}{|c||}{$8$}\\
\hline
\hline
\hline
\multicolumn{5}{||c||}{Downstream $e^\pm$ distribution parameters}\\
\hline
& electrons & positrons & electrons & positrons\\
\hline
f & 1 & 0.6 & 0.92 & 0.8\\
\hline
T' & 25 & 25 & 17 & 16\\
\hline
$\gamma_m$ & - & 100 & 80 & 80\\
\hline
$N_t$ & 0 & $3\times 10^{-3}$ & $1 \times 10^{-3}$ & $2 \times 10^{-3}$\\
\hline
$\alpha$ & - & 1.7 & 2.2 & 1.8\\
\hline
$r$ & - & 4 & 4 & 4\\
\hline
$\gamma_{max}/\gamma_{max}^{th}$& - & 1 & 0.3 & 0.75\\
\hline  
\hline
$\eta$ [\%] & $< 2$ & 20 & 3 & 11\\
\hline
\end{tabular}
\end{center}
\caption{In the upper part of the table we report the values of the paramaters 
with which the simulation was performed: mass ratio between the species; 
number ratio between ions and electrons; fraction of the total upstream 
energy density carried by the ions; ratio between the ions' and electrons' 
energy density upstream. In the lower part of the table, we report the best 
fit values of the parameters in Eq.~\ref{eq:fitdistr}, describing the 
electrons' and positrons' distribution downstream. In the bottom line we report
the energy acceleration efficiency for the pairs, estimated (Eq.~\ref{eq:eta}) 
as the percentage of the total upstream energy density $U_{tot}$ converted into non-thermal $e^\pm$ in each case.}
\label{tab:epp2040}
\end{table}

The results presented in this section are summarized as follows.
When the ions are only 20 times more massive than the pairs and 
40\% of the total number of positive charges, only positron
acceleration is observed.  When a different simulation 
plasma is considered, with the same amount of energy in baryons as before, 
but given to a smaller fraction by number of more massive ions, the fraction of energy
that is converted into accelerated positrons is found to be smaller; their
spectrum is steeper and also the maximum Lorentz factor reached is
lower. At the same time, however, signs of
electron acceleration start appearing, although 
the efficiency of this latter process is lower compared to that of positron
acceleration; the $e^-$ spectrum is softer and their maximum Lorentz factor
smaller than that of $e^+$.   
It is important to notice that while the
results concerning simulation A are in agreement with those of 
\cite{hoshino92}, the simulation B is the first of this kind in the 
literature to show traces of electron acceleration, thanks to the
larger mass ratio and correspondingly lower number ratio between the
ions and the pairs employed. This confirms the role of the
wave polarization in the acceleration process.

\subsection{Relativistic transverse shocks in $e^--e^+-p$ plasmas:
$m_i/m_\pm=100$ \label{sec:high-mass}}

We now analyze how the situation changes increasing the mass-ratio between
the ions and the pairs to $m_i/m_\pm=100$. 

\begin{figure}
\resizebox{\hsize}{!}{
\includegraphics{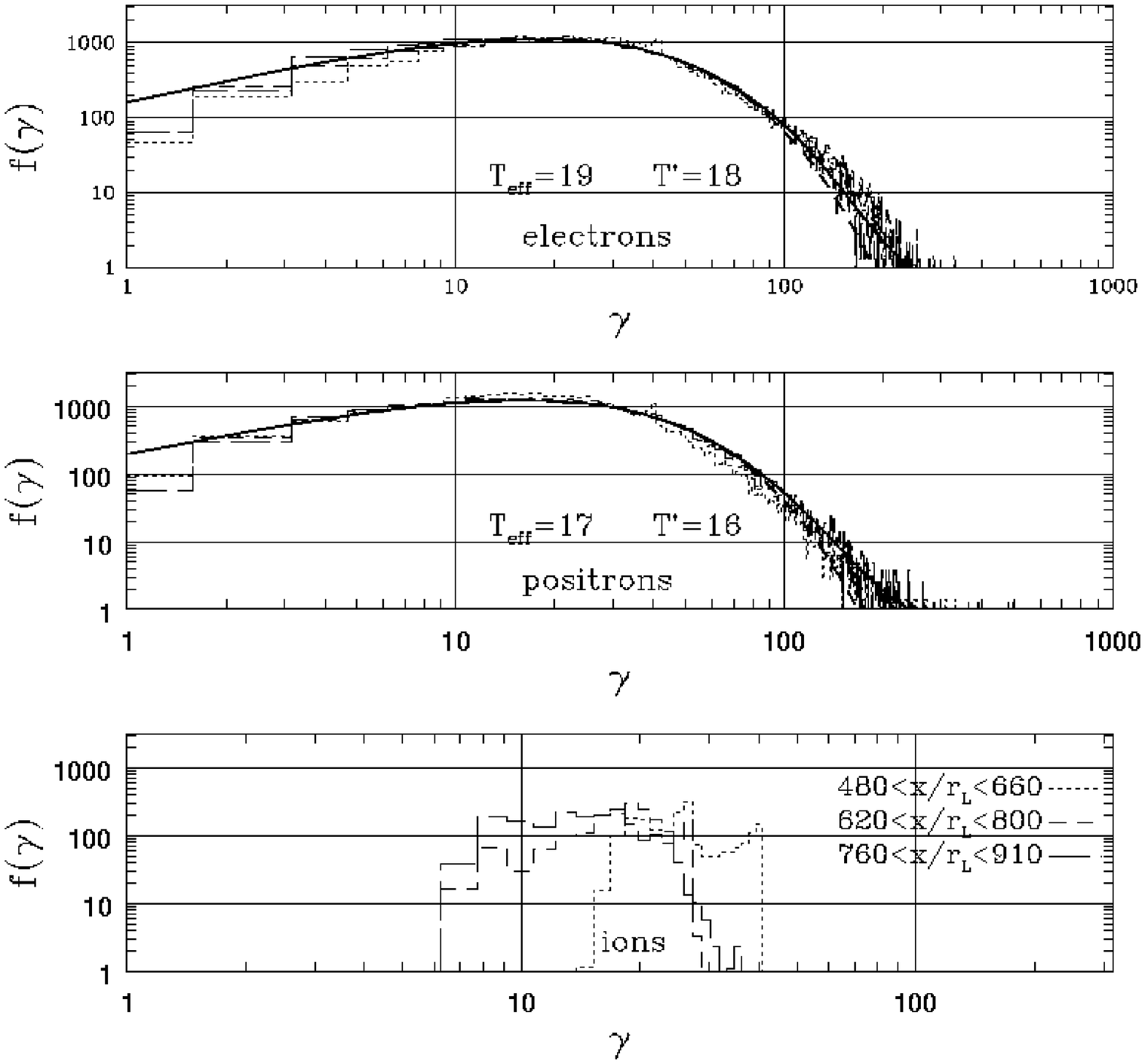}
\includegraphics{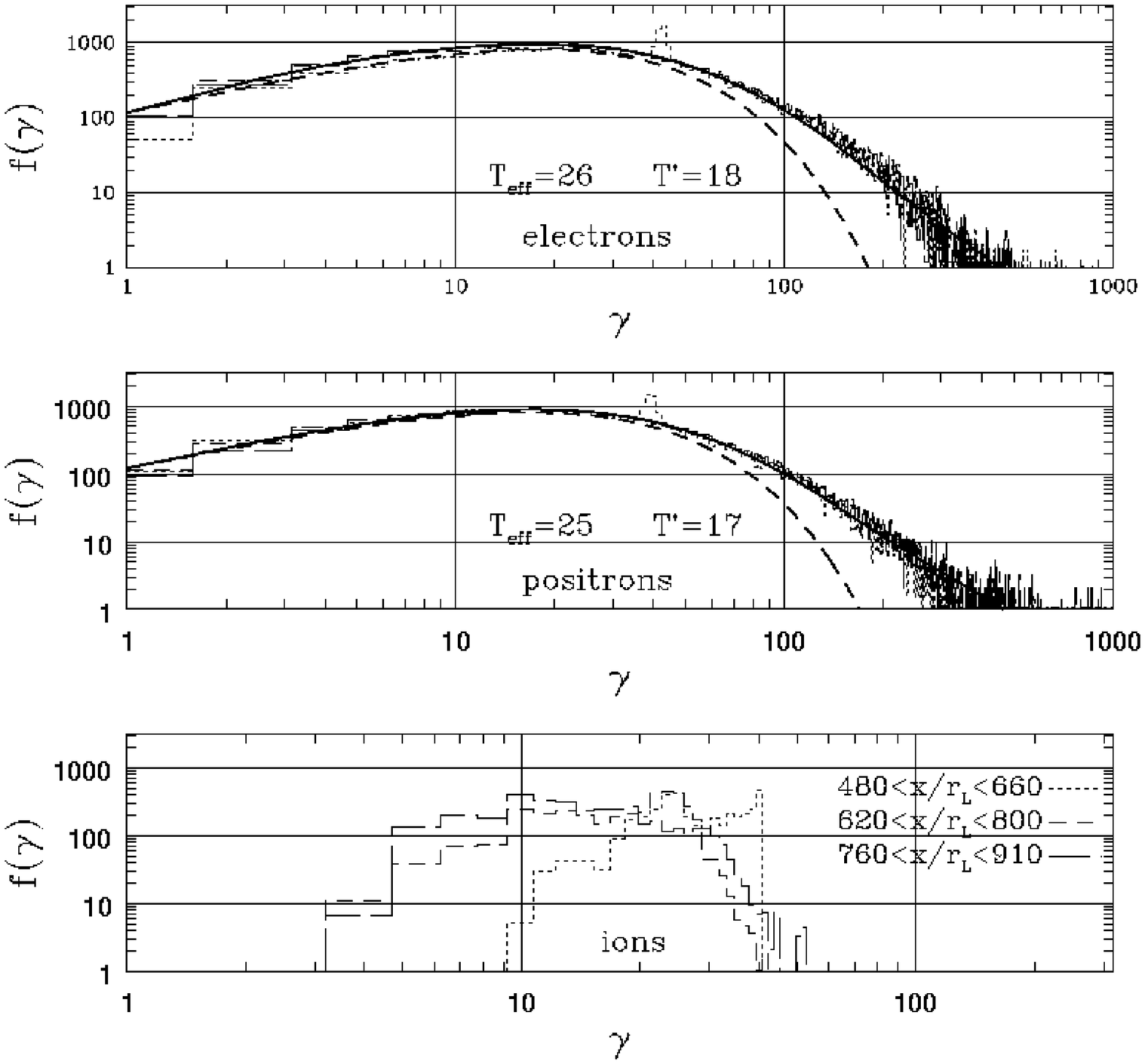}
\includegraphics{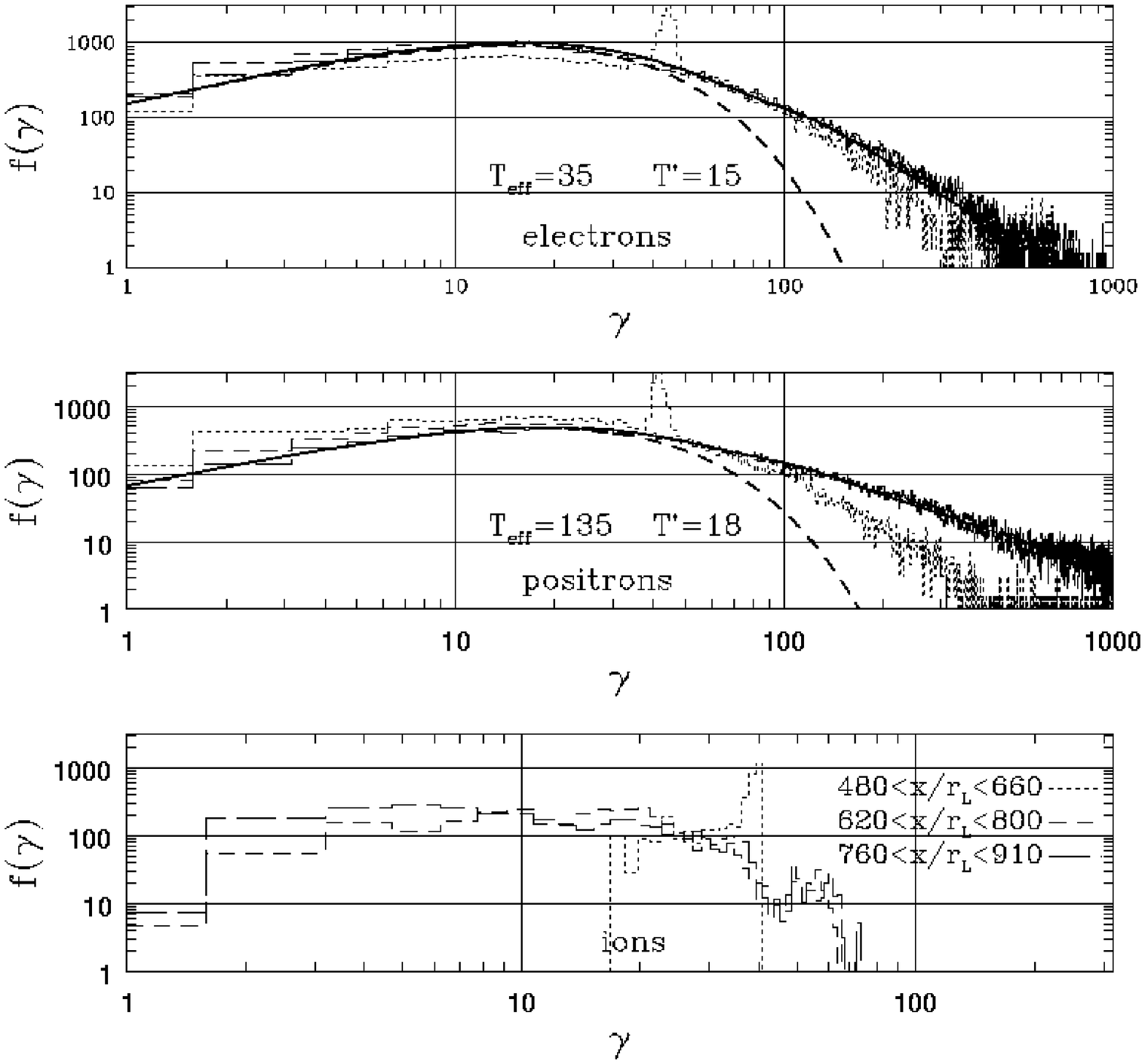}}
\caption{The postshock evolution of the distribution function
of the different species, each normalized to the relative upstream density.
{\bf Left column }: $N_i/N_{-}=0.05$.
{\bf Middle column}: $N_i/N_{-}=0.1$.
{\bf Right column}: $N_i/N_{-}=0.2$.
The notation is the same as for Fig.~\ref{fig:allhists}, with the thick solid
curve representing a fit of the distribution with the function described by
Eq.~\ref{eq:fitdistr}. The best fit values of the parameters are those in 
Table \ref{tab:epp100}.}
\label{fig:allhist100}
\end{figure}

In Fig.~\ref{fig:allhist100} we show the energy distribution of the 
particles at different positions behind the shock, for three simulations
with $m_i/m_\pm=100$ and $N_i/N_{-}$ varying between 0.05 and 0.2. 
In all simulations we employed a box of 10240 cells and the number of 
particles of each species per cell was
12. The size of each cell was $\Delta x=r_{Le}/10$, with $r_{Le}$ being the 
pair Larmor radius upstream. The time-step was $\Delta t=0.9 \Delta x/c$.

\begin{figure}
\resizebox{\hsize}{!}{
\includegraphics{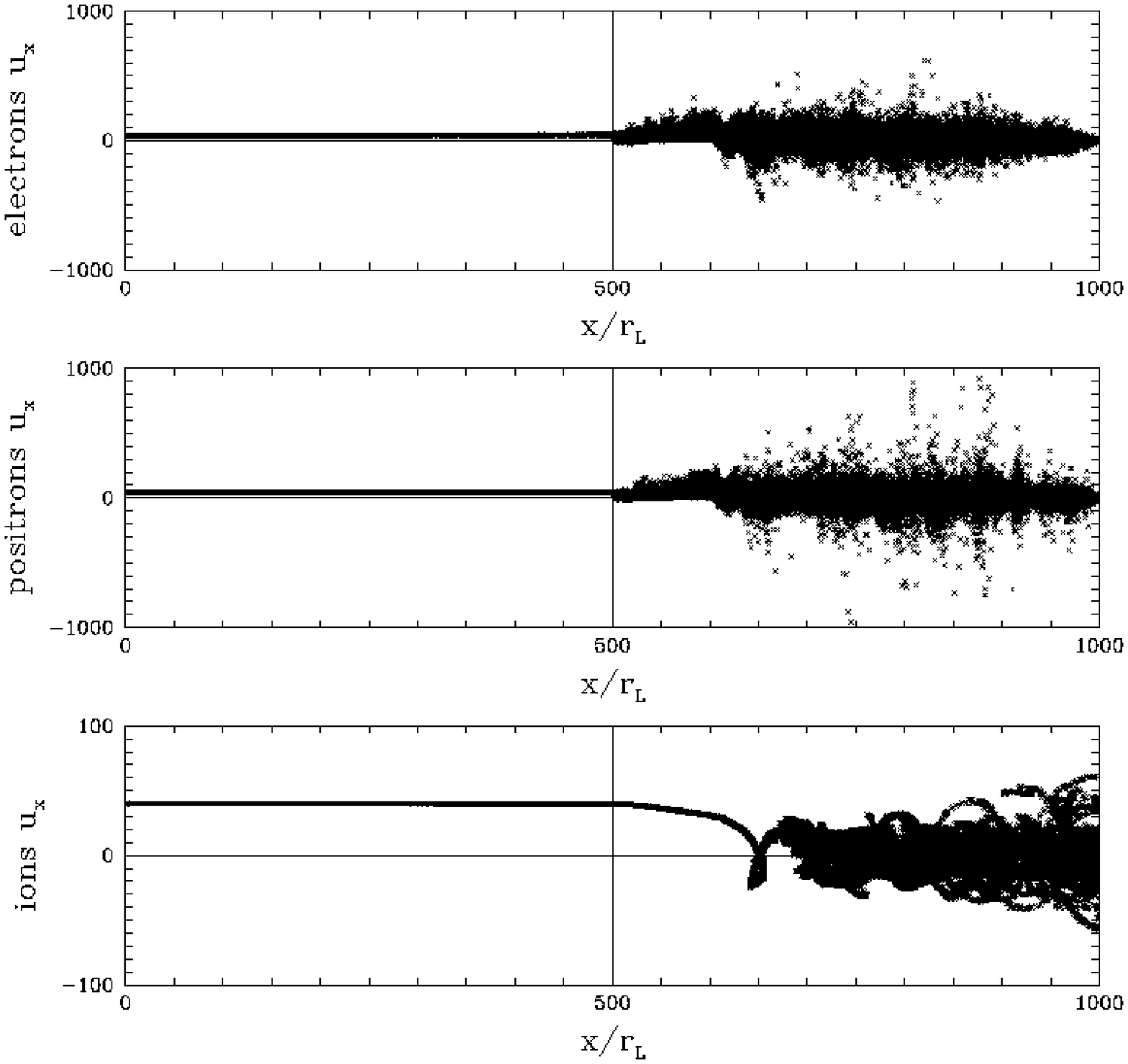}
\includegraphics{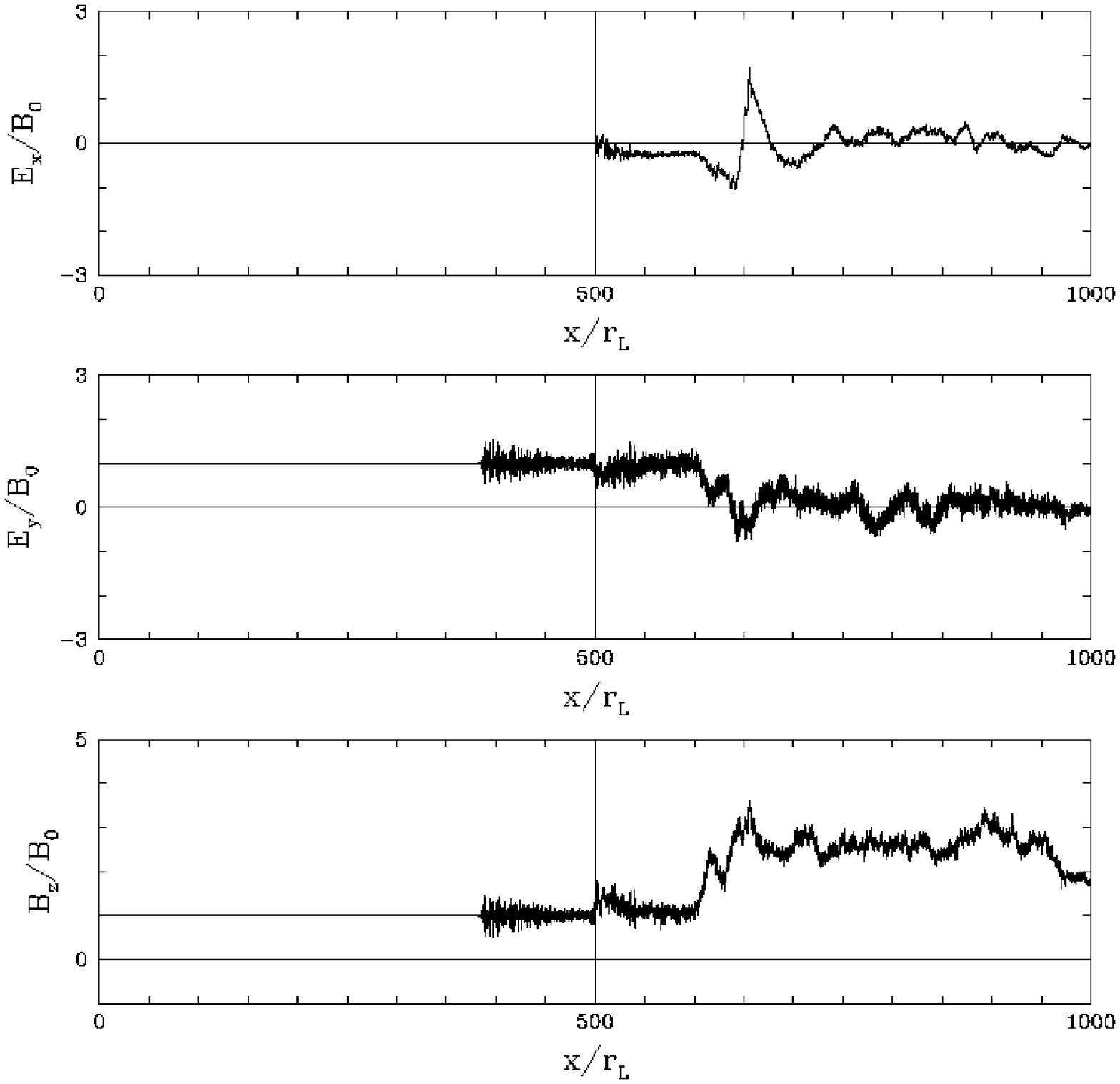}}
\caption{{\bf Left panel:} The ($u_x-x$) section of the 
phase space of the different species for the simulation with $m_i/m_\pm=100$
and $N_i/N_{-}=0.1$.
{\bf Right panel:} The electromagnetic field components, normalized to the
upstream magnetic field intensity, for the same simulation.
In all plots the $x$-coordinate is in units of the pair gyration radius
upstream.}
\label{fig:epp100AB}
\end{figure}
The snapshot of the simulation to which the plots in Fig.~\ref{fig:allhist100}
refer was taken when the shock position was around $x_s/r_{Le}=500$. The 
particle distribution in velocity space and the electromagnetic field 
components
are shown in Fig.~\ref{fig:epp100AB} for the case $N_i/N_{-}=0.1$. 
From comparison of the left panel of this figure, with the left panels
of Figs.~\ref{fig:epp20} and \ref{fig:epp40} one sees that the electrons show
an increasingly large spread in their velocity distribution with
decreasing $N_i/N_{-}$. This feature is made more quantitative by 
Fig.~\ref{fig:allhist100} and Table \ref{tab:epp100}, where simulations 
performed with the same value of the mass ratio ($m_i/m_\pm=100$)
but different fractions of ions are compared.

\begin{table}
\begin{center}
\begin{tabular}{||c||c|c|||c|c|||c|c||}
\hline
\multicolumn{7}{||c||}{Simulation parameters}\\
\hline
\multicolumn{7}{||c||}{$m_i/m_\pm=100$}\\
\hline
$N_i/N_{-}$&\multicolumn{2}{c|||}{$0.05$} &\multicolumn{2}{|c|||}{$0.1$}
&\multicolumn{2}{|c||}{$0.2$}\\
\hline
$U_i/U_{tot}$&\multicolumn{2}{c|||}{$0.56$} &\multicolumn{2}{|c|||}{$0.72$}
&\multicolumn{2}{|c||}{$0.84$}\\
\hline
$U_i/U_{-}$&\multicolumn{2}{c|||}{$5$} &\multicolumn{2}{|c|||}{$10$}
&\multicolumn{2}{|c||}{$20$}\\
\hline
\hline
\hline
\multicolumn{7}{||c||}{Downstream $e^\pm$ distribution parameters}\\
\hline
& electrons & positrons & electrons & positrons & electrons & positrons\\
\hline
f & 0.97 & 0.97 & 0.7 & 0.7 & 0.65 & 0.5\\
\hline
T' & 18 & 16 & 18 & 17 & 15 & 18\\
\hline
$\gamma_m$ & 100 & 100 & 100 & 90 & 100 & 85\\
\hline
$N_t$ & $3.5 \times 10^{-4}$ & $3.7 \times 10^{-4}$ & $2.2 \times 10^{-3}$ & $2.3 \times 10^{-3}$ & $3.2 \times 10^{-3}$ & $4.2 \times 10^{-3}$\\
\hline
$\alpha$ & 3.5 & 3.5 & 3.2 & 2.9 & 2.7 & 1.6\\
\hline
$r$ & 4 & 4 & 3 & 3 & 3 & 3 \\
\hline
$\gamma_{max}/\gamma_{max}^{th}$ & 0.15 & 0.3 & 0.2 & 0.4 & 0.3 & 0.8\\ 
\hline 
\hline
$\eta$ [\%] & $0.9$ & $1$ & $5$ & $5$ & $5$ & $27$\\
\hline
\end{tabular}
\end{center}
\caption{The table refers to simulations employing a mass-ratio of 100 
between the ions and the pairs and number ratios $N_{p1}/N_{e1}$ 
varying between 0.05 and
0.2. In the upper part of the table we report the values of the paramaters 
with which each simulation was performed: 
number ratio between ions and electrons; fraction of the total upstream 
energy density carried by the ions; ratio between the ions' and electrons' 
energy density upstream. In the lower part of the table, we report the best 
fit values of the parameters in Eq.~\ref{eq:fitdistr}, describing the 
electrons' and positrons' distribution downstream. In the bottom line we report
the energy acceleration efficiency for the pairs, estimated (Eq.~\ref{eq:eta}) 
as the percentage of the total upstream energy density $U_{tot}$ converted 
into non-thermal $e^\pm$ in each case.}
\label{tab:epp100}
\end{table}

It is apparent from Fig.~\ref{fig:allhist100} that efficient acceleration of 
both electrons and positrons is found for $N_i/N_{-} \ge 0.1$. A very large
acceleration efficiency is observed in the case when $N_i/N_{-}=0.2$, 
for both
species of particles: 50\% of the positrons and 35\% of the electrons are
found in the suprathermal tail of the distribution in this case, with an 
energy acceleration efficiency that almost reaches 30\% for positrons. 
The ratio
between the efficiencies of electron and positron acceleration is found to
be similar to that in simulation B                        
($m_i/m_\pm=40$, $N_i/N_{-}=0.2$) of the 
previous section, so that the conclusion we derive is that this is strictly
related to the degree of neutrality of the pair plasma, which affects the
polarization of the waves. 

The percentage of energy in suprathermal particles is drastically reduced 
when the number of ions is reduced from $N_i=0.2 N_{-}$ to
$N_i=0.1\ N_{-}$ and as a consequence the energy ratio is halved. 
At the same time, the pair plasma becomes closer to neutral and the behavior 
of the different species of light particles becomes increasingly similar. 
For $N_i = 0.1\ N_{-}$ the energy content of the high energy tails of the 
distributions of the two species is the same, and the 
slopes of the suprathermal power laws become similar, though steeper than 
when the ions are twice as many. 

If $N_i/N_{-}$ is decreased further by a factor of two the acceleration 
efficiency is reduced by a factor of 5 for both species and the power law
indices increase further, becoming exactly the same.

\subsection{Relativistic transverse shocks in $e^--e^+-p$ plasmas: 
Effect of Finite Upstream Temperature \label{sec:temp}}

The results described in \S \ref{sec:dens} and \S \ref{sec:high-mass} 
assumed the upstream flow to
be completely cold, {\it i.e.}, to have zero momentum dispersion. The linear 
theory outlined in
\S \ref{sec:linear} suggests that if the upstream flow is warm enough,
magnetosonic waves at high harmonics of the ion cyclotron
frequency will be absent in the medium behind the shock in the pairs, 
thus reducing the efficiency of injection of pairs into the 
acceleration mechanism. The lack of the harmonics above $n = 
u_1/\Delta u$, where $\Delta u$ is the momentum dispersion of the
upstream ion flow, means that the minimum energy that can be 
accelerated by cyclotron resonance
is  $m_\pm c^2 \gamma_{min} = (m_i c^2 /M_s) \gamma_1$, if 
$M_s = u_1/\Delta u < m_i/m_\pm$.
Thus the suprathermal tail starts farther and farther out on the 
Maxwellian tail of the pairs, for smaller and smaller $M_s$,
thus reducing the acceleration efficiency (and steepening the power 
law part of the spectrum). The
maximum energy possible is still set by resonance with the 
fundamental, but now there are many fewer
such particles.

The simulations whose results are shown in the following were performed
assuming gaussian functions to describe the particles' distribution upstream 
of the shock:
\begin{equation}
f(u_x,u_y)=\frac{1}{\pi \Delta u^2}
\exp\left[-\frac{(u_x-u_1)^2}{\Delta u^2}\right]\ 
\exp\left[-\left(\frac{u_y}{\Delta u}\right)^2\right] .
\label{eq:tsdistr}
\end{equation}
$M_s \equiv u_1/\Delta u$ is the relativistic flow sonic Mach number 
as defined in the 
linear instability theory.

All three species of particles were injected in the box according to 
the distribution function in Eq.~\ref{eq:tsdistr}, with $u_1$ corresponding
to a Lorentz factor $\gamma_1=40$ and $\Delta u =0.1\ u_1$. The mass ratio
between the ions and the pairs was taken to be $m_i/m_\pm=100$ and two
different simulations were performed, corresponding to $N_i/N_{-}=0.1$ and
$N_i/N_{-}=0.2$. Except for the substitution of a Dirac $\delta$ with a
gaussian as the function describing the particles distribution, the
simulation set-up is exactly the same as that described in \S\S 
\ref{sec:high-mass},
{\it i.e.} same number of cells and of particles per cell, same cell-size
(in units of the {\em mean} pair Larmor radius upstream) and time-step.

\begin{figure}[H]
\resizebox{\hsize}{!}{
\includegraphics{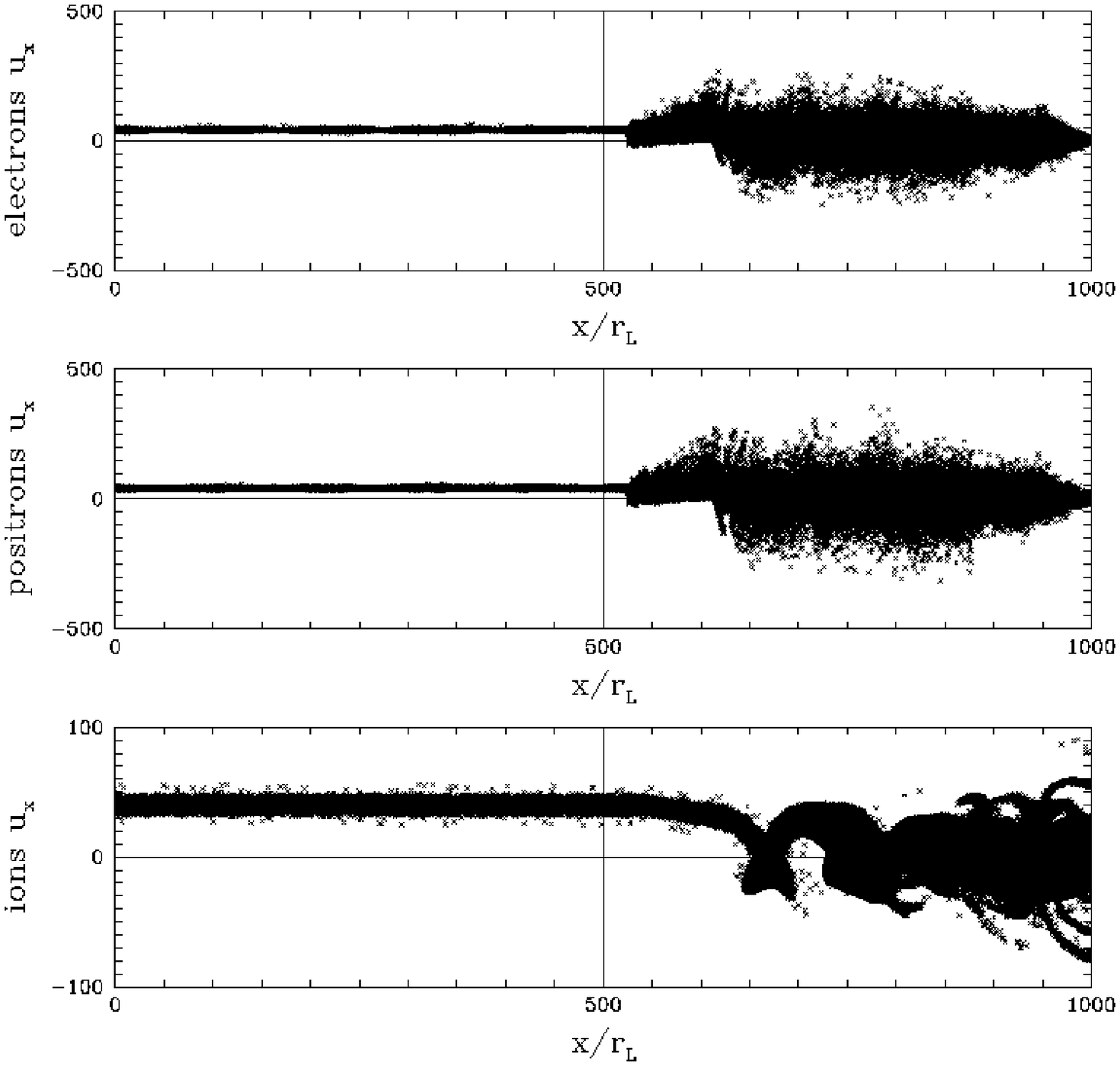}
\includegraphics{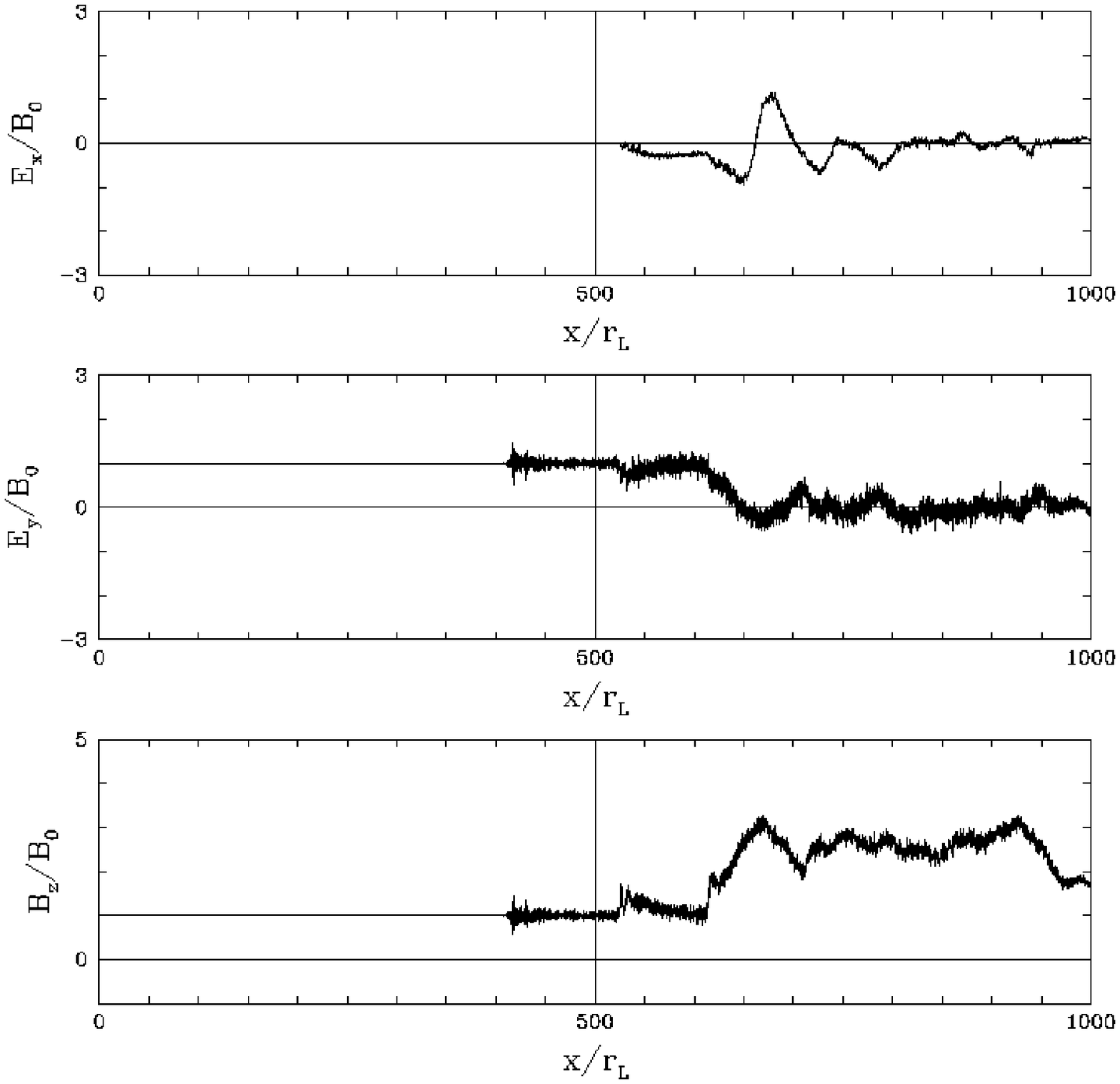}}
\caption{{\bf Left panel:} The ($u_x-x$) section of the 
phase space of the different species for a simulation with $m_i/m_\pm=100$
and $N_i/N_{-}=0.1$. The particle distribution function upstream of the
shock is described by Eq.~\ref{eq:tsdistr} with $u_1 \simeq 40$ and 
$\Delta u =0.1\ u_1 \; (M_s = u_1/\Delta u = 10)$. The shock front is at 
$x/r_L \simeq 500$.
{\bf Right panel:} The electromagnetic field components, normalized to the
upstream magnetic field intensity, for the same simulation.
In all plots the $x$-coordinate is in units of the {\em mean} 
(corresponding to $\gamma_1$) pair gyration radius upstream.}
\label{fig:epp100ABT}
\end{figure}

Fig.~\ref{fig:epp100ABT} is a snap-shot of the simulation
with $N_i/N_{-}=0.1$, taken when the shock front is at $x/r_L \simeq 500$.
In this figure we plot the particles' phase-space and the electromagnetic 
fields in the simulation for a direct comparison with those in 
Fig.~\ref{fig:epp100ABT}, which referred to the same incoming plasma 
taken to be completely cold.  

From the comparison of Fig.~\ref{fig:epp100ABT} with Fig.~\ref{fig:epp100AB} 
it is immediately
apparent that the maximum pair energy downstream is reduced when a spread
in the particle velocities upstream is introduced.

\begin{figure}[H]
\resizebox{\hsize}{!}{
\includegraphics{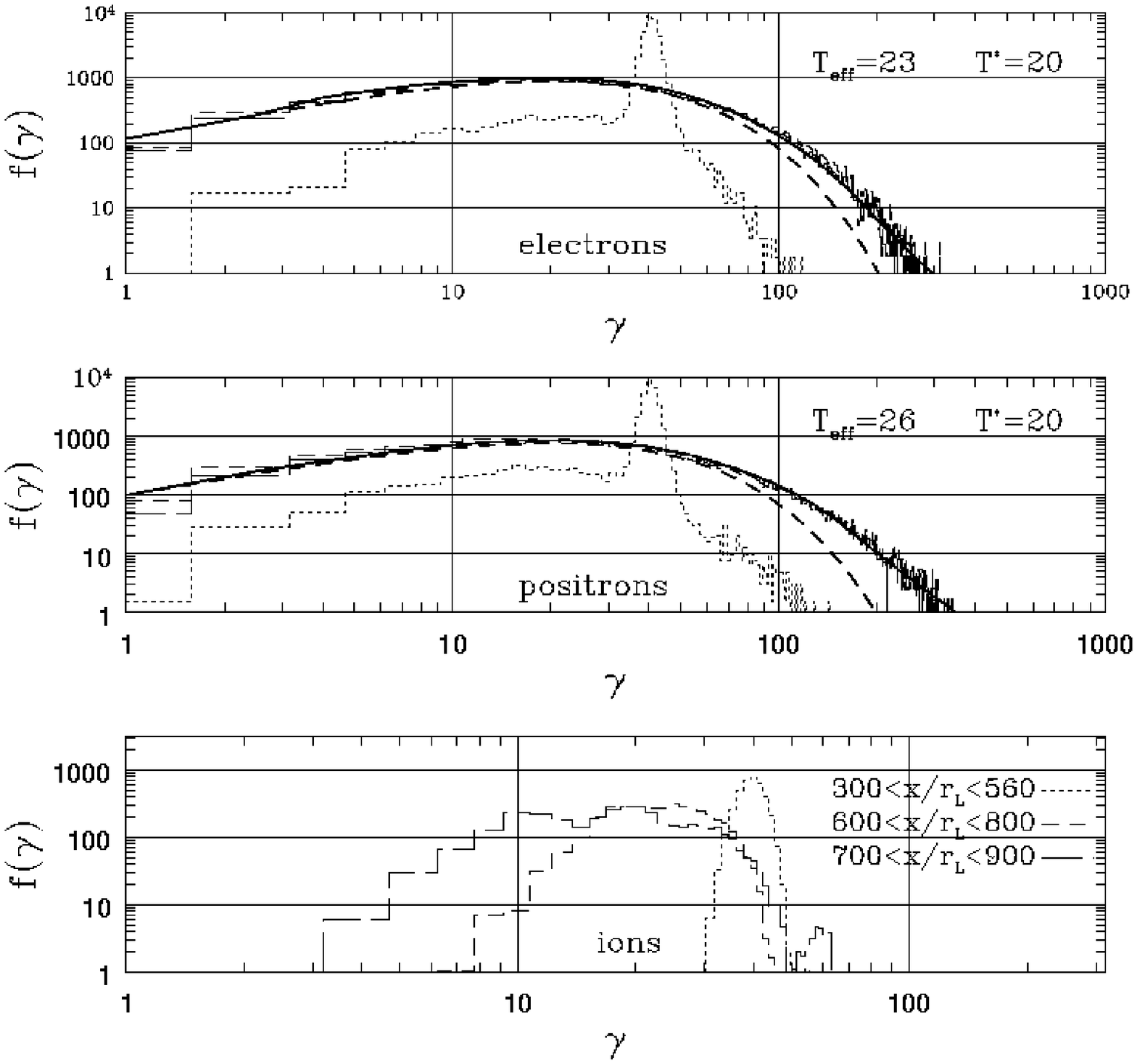}
\includegraphics{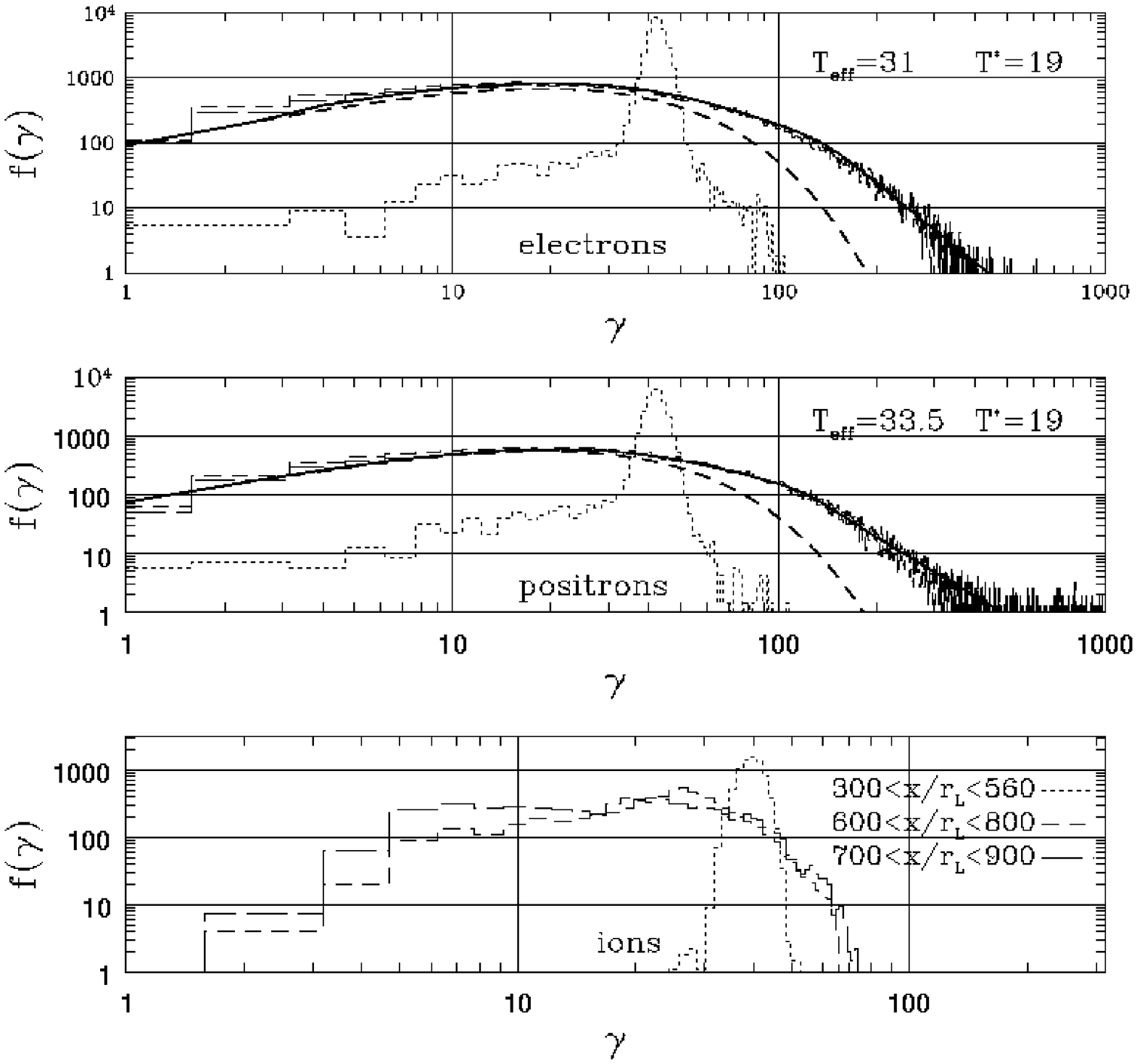}}
\caption{The postshock evolution of the distribution function
of the different species, each normalized to the relative upstream density.
{\bf Left column }: $N_i/N_{-}=0.1$.
{\bf Right column}: $N_i/N_{-}=0.2$.
The notation is the same as for Fig.~\ref{fig:allhist100}, with the thick
solid curve representing a fit of the distribution with the function 
described by Eq.~\ref{eq:fitdistr}. The best fit values of the parameters 
are those in 
Table \ref{tab:epp100T}.}
\label{fig:tsallhist}
\end{figure}

The differences are even more evident from the comparison between 
Fig.~\ref{fig:allhist100} and Fig.~\ref{fig:tsallhist}. In the latter 
we plot the particle 
distribution functions as extracted from slices of the simulation at 
different distances from the shock front. The left and right panels 
refer to the case when $N_i/N_{-}=0.1$ and $N_i/N_{-}=0.2$ respectively, 
and hence should
be compared to the middle and right panel in Fig.~\ref{fig:allhist100}. 
The notation is the same as for that figure.

\begin{table}
\begin{center}
\begin{tabular}{||c||c|c|||c|c||}
\hline
\multicolumn{5}{||c||}{Simulation parameters}\\
\hline
&\multicolumn{2}{c}{$m_i/m_\pm=100$}&  \multicolumn{2}{c||}{$\Delta u/u_0=0.1$}\\
\hline
\hline
$N_i/N_{-}$&\multicolumn{2}{|c|||}{$0.1$}
&\multicolumn{2}{|c||}{$0.2$}\\
\hline
$U_i/U_{tot}$&\multicolumn{2}{|c|||}{$0.72$}
&\multicolumn{2}{|c||}{$0.84$}\\
\hline
$U_i/U_{-}$&\multicolumn{2}{|c|||}{$10$}
&\multicolumn{2}{|c||}{$20$}\\
\hline
\hline
\hline
\multicolumn{5}{||c||}{Downstream $e^\pm$ distribution parameters}\\
\hline
& electrons & positrons & electrons & positrons\\
\hline
f & 0.85 & 0.8 & 0.6 & 0.6\\
\hline
T' & 20 & 20 & 19 & 19\\
\hline
$\gamma_m$ & 120 & 110 & 120 & 100\\
\hline
$N_t$ & $10^{-3}$ & $2 \times 10^{-3}$ & $3 \times 10^{-3}$ & $4 \times 10^{-3}$\\
\hline
$\alpha$ & 4.5 & 4. & 4. & 3.3\\
\hline
$r$ & 4 & 4 & 4 & 4 \\
\hline
$\gamma_{max}/\gamma_{max}^{th}$ & 0.09 & 0.1 & 0.15 & 0.65 \\
\hline
\hline  
$\eta$ [\%] & 2.5 & 4 & 4 & 4\\
\hline
\end{tabular}
\end{center}
\caption{The table refers to simulations employing a mass-ratio of 100 
between the ions and the pairs and a finite temperature upstream:
$\Delta u/u_0=0.1$ (plots in Fig.~\ref{fig:tsallhist}). 
In the upper part of the table we report the values of the paramaters 
with which each simulation was performed: 
number ratio between ions and electrons; fraction of the total upstream 
energy density carried by the ions; ratio between the ions' and electrons' 
energy density upstream. In the lower part of the table, we report the best 
fit values of the parameters in Eq.~\ref{eq:fitdistr}, describing the 
electrons' and positrons' distribution downstream. In the bottom line we report
the energy acceleration efficiency for the pairs, estimated (Eq.~\ref{eq:eta}) 
as the percentage of the total upstream energy density $U_{tot}$ converted 
into non-thermal $e^\pm$ in each case.}
\label{tab:epp100T}
\end{table}

The particle acceleration efficiency is  reduced
with respect to the case when the upstream plasma was taken to be completely
cold. As one can see by comparing the values of the parameters in
Table \ref{tab:epp100T} with those in Table \ref{tab:epp100}, for both 
values of $N_i/N_{-}$,
the fraction of energy carried by suprathermal particles is smaller and 
the power-law index of the suprathermal tail is steeper.

The resonance condition for absorption of an ion cyclotron wave by an electron or
positron requires that $n \Omega_{ci} = \Omega_{c\pm} $. Thermal 
dispersion $\Delta u /u_1 > m_\pm /m_p$ in the upstream
ions prevents growth for harmonic number $n > M_s = u_1 / \Delta u$, 
which cuts the efficiency
of generation of a non-thermal tail to the pairs' momentum distributions.

\section{Discussion and Conclusions \label{sec:discuss}}

In accord with previous results (\cite{langdon88,gallant92}), 
we find full thermalization in pure $e^- - e^+$ shocks.  The reason
is the rapid reabsorption of waves at and above the upper hybrid frequency
$\Omega_{UH\pm}$, which causes the shock to behave as if it were collisional 
- rapid emission of the EM waves by the relativistic cyclotron instability at 
the particle reflection front, with rapid reabsorption of the waves, and 
exchange of real photons, is a mechanism just as capable as the exchange of 
virtual photons in Coulomb collisions in producing local thermodynamic 
equilibrium.  The conclusion, that pure pair shocks yield relativistic
Maxwellians in the downstream medium, is not limited to the one-dimensionality 
of our model, as is 
shown elsewhere in 3D simulations by Spitkovsky and Arons (in preparation). Those
simulations show that at sufficiently low $\sigma$ ($\sigma < 10^{-3}$ is low
enough), the relativistic Weibel instability ({\it e.g.}, \cite{yoon87}) takes over from 
cyclotron instability in the pairs as the deceleration and thermalization mechanism
of the shock wave, even when the background magnetic field is fully transverse
to the flow.  The physical reasons for this result are discussed in detail elsewhere,
in the context of the 3D results\footnote{Other recent simulations of Weibel
mediated shocks in $e^\pm$ and $e-p$ plasmas (\cite{silva03, hed04, fred04, hed05, jaro05})
have had insufficient size and temporal length (or have been constrained by periodic
boundary conditions in the flow direction) to show the development of the
shock wave; as a result, they have been, in effect, studies of the foreshock.}.  
However, the essence is that the magnetic fields generated by the Weibel instability
in an unmagnetized shock saturate in a tangled network of tangled 
magnetic filaments (\cite{milos06}), taking a distance $l_{therm} \sim 25 c/\omega_{p1}$.  
These filaments scatter and thermalize the 
incoming particles.  Since $l_{therm} /r_{L1\pm} \sim 25 \sqrt{\sigma_{1\pm}}$, Weibel
based thermalization (a multi-diemensional effect) controls the shock structure for
$\sigma_{1\pm} < 0.002$.  For larger magnetizations, the cyclotron effects studied here
dominate, somewhat modified by mild crinkling of the shock front. 

In the shock wave, the incoming flow 
continuously injects cold ($M_s \gg 1$) ions into the pair plasma, the latter 
being already heated by the leading shock in the pairs alone. These freshly 
injected cold ions efficiently produce high harmonic 
waves which are cyclotron resonant with the pairs, even though 
the heated ions that are found further downstream from the pair shock do not 
efficiently radiate high harmonic waves. Magnetic reflection at the shock 
front produces a continuous supply of cold gyrating ions, thus ensuring  
energy transfer from the ions to the non-thermal pairs 
in the downstream medium, different from what is the case in the analogous 
initial value problem in a uniform medium. 

When the ion density is a small fraction of the pair density, non-thermal 
acceleration becomes symmetric between electrons and positrons. It also 
becomes less efficient, leading to steeper power-law slopes in the 
non-thermal part of the spectrum. As shown in Fig.~\ref{fig:allhist100}, 
the spectra take a Maxwellian form up to energies around 
$E_\pm \approx \gamma_1 m_\pm c^2$, then have a power law form at higher 
energies. The energy above which the power law distribution sets in, and 
the slope, depend on $N_i / N_\pm$.  Thus, apparent universality (if it exists) of power 
law spectra in synchrotron sources where {\it relativistic} shocks are 
inferred to play a role in non-thermal accleration would be
a consequence of universality in the composition of the upstream flow, 
expressed as the ratio $N_{1i}/N_{1\pm} $, if the cyclotron resonance 
mechanism described here really is the dominant non-thermal heating process 
and if
all relativistic shocks are sufficiently close to the transverse
geometry studied here. 

At present, nothing has been published on the full
3D structure of these shocks\footnote{\cite{hed04} in essence study the foreshock
of an $e-p$ relativistic shock in 3D with a mass ratio $m_p /m_e = 15$.}.
Our expectation is that the 3D structure of electron-positron-ion 
transverse shocks will show a transition to unmagnetized physics at a
low value of $\sigma_{1,total}$.  The value of this transition magnetization
is quite uncertain, since current neutralization of the ion filaments
formed by the Weibel instability by electron return currents (\cite{lyub06})
may greatly increase the thermalization length, thus giving more room in
$\sigma$ space for cyclotron dynamics to ultimately mediate the shock structure.
These issues are under investigation.

Relativistic flows that get 
their energization from Poynting losses from compact objects (possibly 
all in our list of pulsar winds, microquasar, AGN and GRB jets) may 
well be in this category, since the pair and ion content may both be set 
by the underlying electrodynamics of the compact objects (rotating neutron 
star, magnetized disk.) Quantification of how close shocks have to be to 
transverse requires consideration of a wider range of shock geometries.
Investigation of other non-thermal heating mechanisms, such as Diffusive Shock
Acceleration and magnetic pumping, require higher dimensional simulations.
Spitkovsky and Arons (in preparation) address both of these issues.

In the shocks, the relative proportion of electron versus positron 
acceleration follows the ion to pair density, as expected from the 
changes in wave polarization inferred from the linear theory.   

Our study of the shock acceleration process confirms, therefore, 
the results found by \cite{hoshino92}, extending them to a
higher mass ratio. Also confirmed is the expectation that signs of
electron acceleration would be found if one could consider a plasma
that contained the same amount of upstream flow energy given to a 
smaller fraction 
of ions. The overall evolution of the particles is consistent with resonant 
cyclotron absorption of magnetosonic waves by the pairs, whose frequencies 
are harmonics of the ion cyclotron frequency, being the main process 
providing non-thermal acceleration. Magnetic reflection of the upstream 
ions in the shock front sets up the unstable ion cyclotron motion which causes
the wave emission. The non-thermal acceleration efficiencies and power law 
indices of the particle spectra are shown, as a function of the basic 
governing parameter $N_i/N_{-}$, in Table \ref{tab:epp100}, 
when the upstream flow is cold. The reduction of the acceleration 
efficiency for a single value of the Mach number 
$M_s = 10 \ll m_i/m_\pm = 100$, with results summarized in 
Table \ref{tab:epp100T}, shows the strong effect of large upstream momentum 
dispersion in the upstream flow. Equivalently, the absorption of ion cyclotron
waves in the shock structure by the pairs is a strong non-thermal acceleration
mechanism when the
upstream flow is quite cold ($M_s > m_i/m_\pm $), as results from adiabatic 
expansion in relativistic outflows from compact objects.

We conclude by pointing out an interesting feature of this acceleration 
mechanism. The acceleration efficiency and the spectral form of the 
accelerated 
particles depend on the composition of the flow upstream of the shock. 
That composition depends on the source of the relativistic flow
within which the shock wave occurs. Relativistic flows of the type 
considered here are thought
to be the consequence of magnetized wind and jet acceleration by 
compact objects, and pair
creation within and near such objects. These properties are the 
result of the compact objects'
electrodynamics, which constrains some aspects of the outflowing 
plasma composition to
particular values - for example, in pulsars and magnetars, the ion 
flux is constrained to be the
Goldreich-Julian value.
Thus, one might expect a correlation between the radiative 
emission properties of the shocked flow and the emission properties of the 
underlying compact object, as appears to be the case in the X-ray emission 
from  pulsar wind nebulae and the spectra of the pulsed X-ray emission from 
the pulsars driving those nebulae (\cite{gotthelf03}).

\acknowledgments
J.A. was assisted by financial support from NASA grants NAG5-12031, TM4-5005X
and NNG06G108G, and by NSF grant AST-0507813, all awarded to the University 
of California, Berkeley.
He also benefited from the support of the Miller Institute for Basic 
Research in Science, and from the forbearance of the taxpayers of California. 
We accomplished part of the work while we enjoyed the hospitality of the 
Kavli Institute for Theoretical Physics in April 2005.

\appendix
\section{Relativistic Cyclotron Instability of Ring Distributions in a 
Uniform Magnetized \epp\ plasma \label{sec:ring}}

We consider an overall neutral \epp\ plasma uniformly distributed in space and
permeated by a uniform magnetic field. The initial distribution of the plasma
in momentum space is described as rings, with all species gyrating in
the background magnetic field with the same Lorentz factor, 
$\gamma_{+}=\gamma_{-}=\gamma_{p}$. For most of the work in this section,
we assume that all the species are cold, that is, have no dispersion around their gyration 
momenta. In preparation for the shock studies described in \S\ref{sec:shocks}, we also
briefly describe some results of the linear instability theory when the ions gyrate 
with a small momentum dispersion in a relativistically hot pair plasma.

Cold ring distributions are known
to be unstable to cyclotron emission and the growth-rate of the instability
scales proportionally to the particles' cyclotron frequency.
Hence, in the situation we are presently considering the
process of relaxation of the particles' distribution functions will
show two different stages. 

Initially, the pairs, whose gyration frequencies are much larger than 
those of the ions, will undergo cyclotron instability and tend to thermalize
through emission and absorption of relativistic cyclotron waves.
During this stage, the ions form a static uniform neutralizing
background: they cannot interact efficiently with the waves produced
by the pairs, which are too high frequency, and therefore their distribution
is expected to be left nearly unchanged while the pairs relax.
Only at later times the ions will develop the same instability, with the 
delay determined by the difference
in the Larmor frequencies of the lighter and heavier particles.

In the case of a plasma with about equal amount of energy carried by 
light particles and magnetic fields 
($\sigma_{-}\simeq\sigma_{+}\simeq 1$), the time-scale for the cyclotron 
instability in the pairs to reach saturation turns out to be of
order a few tens of gyration periods, while it is shorter for lower
magnetization (see \cite{hoshino91}).
Hence, if one considers, as we shall do in the following, a plasma 
with magnetization $\sigma\lessim 1$,
for high enough mass ratios, the expectation is that the instability of
the pairs will have reached saturation, and they will have already almost 
completely thermalized, before the ion cyclotron instability sets in. 

\subsection{Linear theory of the ion cyclotron instability \label{sec:linear}}
The ion distribution function is described
by:
\begin{equation}
f_i(u_\perp, u_\parallel)={1 \over 2 \pi u_{0}} \delta(u_\perp-u_{0})
\delta(u_\parallel),
\label{eq:ring}
\end{equation}
where $u$ is the particle four-velocity ($u=\gamma \beta$, with 
$\gamma=1/\sqrt{1-\beta^2}$ being the particle Lorentz factor),
$u_{0}$ is the initial value of $u$ and the symbols $\parallel$ 
and $\perp$ are with respect to the magnetic field direction.
We assume the magnetic field aligned
with the z-axis of a cartesian system of coordinates ${\bf B}=B {\bf e}_z$.
Since we are interested in the particles' cyclotron emission the waves we
are going to consider will have both the electric field and wave vector
perpendicular to the magnetic field direction (extraordinary mode).
If we assume 
${\bf k} = k {\bf e}_x$ the dispersion relation can be written as:
\begin{equation}
{c^2 k^2 \over \omega^2}=\varepsilon_{yy}-{\varepsilon_{xy} \varepsilon_{yx} \over \varepsilon_{xx}}\ .
\label{eq:gendisp}
\end{equation}

The general expression for the dielectric tensor in a uniform magnetized 
plasma can be found, for example, following the theory
in the book by \cite{krall&tri} (see also \cite{hoshino91}).
We mentioned above that when the ion instability is still in its early 
phases, the pairs will have already exhausted their free energy. This 
fact leads us to expect the fluid approximation to appropriately describe 
the pairs' contribution to the propagation characteristics of the waves 
whose growth is stimulated by the ions. We will therefore describe the
pairs' response as that of a relativistically 
hot fluid with effective temperature $T_\pm$ ($k_B T_\pm=P_\pm/N_\pm$ where 
$P_\pm$ is the pressure and $N_\pm$ the number density) and adiabatic 
index $\Gamma$. The latter would be $\Gamma=4/3$ for a 3D relativistic plasma
and $\Gamma=3/2$ for a 2D relativistic plasma: we confine our attention to 
the 2D case, since our spatially 1D simulations do not heat the plasma in 
the magnetic field direction. 

The relevant components of the dielectric tensor will then be given by:
$$
\varepsilon_{xx} =
1+2 \pi \sum_{n=-\infty}^\infty {n^2 \over \nu_i^2} {\gamma_0^2 \over 
\sigma_i \alpha_i^2} \int du_\parallel du_\perp { \partial f_i \over  \partial u_\perp}\ 
{(J_n (\alpha_i u_\perp / \gamma_0))^2 \over (\sqrt{1+u_\perp^2}/
\gamma_0-n/\nu_i)}+
$$
\begin{equation}
\label{eq:etotxx}
+\left({1 \over \sigma_{-}} + {1 \over \sigma_{+}} \right) {1 \over 
(1-\nu_e^2 + (\Gamma-1)\alpha_e^2)}\ ,
\end{equation}
$$
\varepsilon_{xy} = -\varepsilon_{yx} =
2 \pi {\it i} \sum_{n=-\infty}^\infty {n \over \nu_i^2} {\gamma_0 \over 
\sigma_i \alpha_i} \int du_\parallel du_\perp u_\perp { \partial f_i \over  \partial 
u_\perp}\ {J_n (\alpha_i u_\perp / \gamma_0)\ J_n' (\alpha_i u_\perp / 
\gamma_0) \over (\sqrt{1+u_\perp^2}/\gamma_0-n/\nu_i)}+
$$
\begin{equation}
\label{eq:etotxy}
+{{\it i} \over \nu_e} \left( {1 \over \sigma_{+}} - {1 \over \sigma_{-}} 
\right) {1 \over (1-\nu_e^2 + (\Gamma-1)\alpha_e^2)}\ ,
\end{equation}

$$
\varepsilon_{yy} =
1+2 \pi \sum_{n=-\infty}^\infty {1 \over \nu_i^2} {1 \over \sigma_i} \int 
du_\parallel du_\perp u_\perp^2 { \partial f_i \over  \partial u_\perp}\ {(J_n' 
(\alpha_i u_\perp / \gamma_0))^2 \over (\sqrt{1+u_\perp^2}/\gamma_0-
n/\nu_i)}+
$$
\begin{equation}
\label{eq:etotyy}
+\left({1 \over \sigma_{-}} + {1 \over \sigma_{+}} \right) 
{1-(\Gamma-1) \alpha_e^2/\nu_e^2 \over 
(1-\nu_e^2 + (\Gamma-1)\alpha_e^2)}\ ,
\end{equation}
where we have used the following definitions:
\begin{equation}
\label{eq:sigalfnu}
\sigma_{s}={\Omega_{c0s}^2 \over \Omega_{p0s}^2} = {B_0^2 \over 4\pi m_s N_s \gamma_{0s} c^2}\ , \, \, \, \,  
\alpha_s={ k c \over \Omega_{c0s}}\ , \, \, \, \,  \nu_s={\omega \over \Omega_{c0s}}\ ,
\end{equation}
with 
\begin{eqnarray}
\label{eq:opc}
\Omega_{p0s}=\sqrt{\frac{4 \pi N_s e_s^2}{m_s \gamma_{0s}}} = {\omega_{ps} \over \sqrt{\gamma_{0s}}}
& {\rm and} & 
\Omega_{c0s}=\frac{ e_s B_0 }{ m_s \gamma_{0s} c} = {\omega_{cs} \over \gamma_{0s}} ,
\end{eqnarray}
the plasma and cyclotron frequencies of particles of species $s$ respectively.
The functions $J_n$ that appear in the Eqs.~\ref{eq:etotxx}-\ref{eq:etotyy} are
the ordinary Bessel functions of the first kind of index $n$ and argument
$z=c \beta_i k/\Omega_{ci}$.

Moreover, here, $\gamma_{0i}$ is equal to the initial Lorentz factor of the
plasma $\gamma_0$, whereas $\gamma_{0\pm}$, entering the quantities that refer 
to the pairs,
is defined by the temperature of the pairs
after the saturation of their instability: $\gamma_{0\pm}=k_B T_\pm/m_\pm c^2$. 
The relation $\gamma_{0\pm}< \gamma_0$ is going to hold, since, even if the 
fraction of energy
the pairs lost to waves during the instability were negligible, the
pressure of a Dirac $\delta$ distribution function centered on 
$\bar \gamma$ corresponds to that of a maxwellian with 
$\gamma=(\bar \gamma-1)/2 \simeq \bar \gamma/2$. 

If the ions can be treated as a cold fluid when the
instability sets in for them, one can perform the integrations in 
Eqs.~\ref{eq:etotxx}-\ref{eq:etotyy} by parts 
(see \cite{hoshino91}). 

\begin{figure}
\resizebox{\hsize}{!}{
\includegraphics{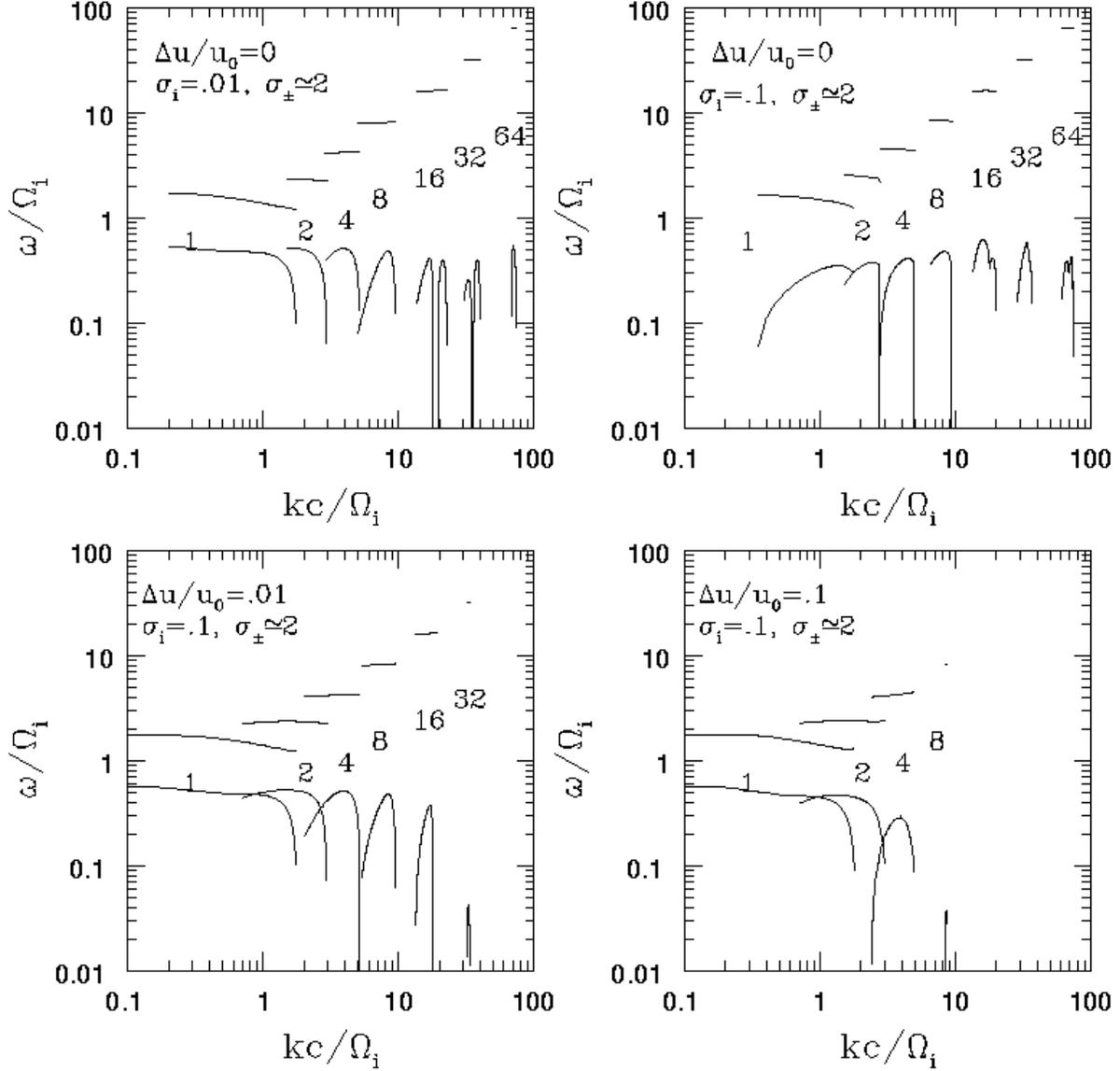}}
\caption{The dispersion relation for waves produced by the ion 
cyclotron instability. A mass ratio of $2000$ between the ions 
and the pairs is 
used, and the energy balance between the different components corresponds to 
$\sigma_i=0.1$ and $\sigma_{-}=2 \simeq \sigma_{+}$.
{\bf Upper left panel}: the dispersion equation is solved in the approximation 
that the ions are completely cold and that mode coupling is negligible 
($\varepsilon_{xy}= 0$). 
{\bf Upper right panel}: ions treated as cold but mode coupling included. 
{\bf Lower left panel}: a thermal spread of the ions corresponding to 
$\Delta u/u_0=0.01$ (see text) is assumed.
{\bf Lower right panel}: a thermal spread of the ions corresponding to
$\Delta u/u_0=0.1$ (see text) is assumed. While in the upper panels both the 
electromagnetic and magnetosonic wave branches have non-negligible
maximum growth-rates, only magnetosonic waves are found to grow in the 
presence of a finite spread in the temperature of the ions.}
\label{fig:disprel}
\end{figure}

In general, solutions of the dispersion relation are found numerically and in 
Fig.~\ref{fig:disprel} we plot the results in a few cases relevant for
the present study. In all plots we consider a plasma with $m_i/m_\pm=2000$ 
and with a baryon fraction $N_i/N_{-}=0.01$. 
The two upper panels refer to the case when the ions can be treated as
completely cold (analogous to \cite{hoshino91}). 
In the upper left panel the pair plasma
is approximated as neutral and the coupling between the longitudinal and 
transverse components of the extraordinary mode is neglected 
$\varepsilon_{xy} = 0$. In the upper right panel roots of the complete 
dispersion equation are shown.

When the coupling is neglected, at high harmonic numbers, the dispersion
relation shows two separate unstable branches, corresponding to 
electromagnetic ($\omega \approx k c$) and
magnetosonic waves ($\omega \approx k c \sqrt{(1/2+\sigma_\pm)/(1+\sigma_\pm)}$). 
The growth-rate of magnetosonic waves is  more or less constant with increasing $n$ while the growth of
electromagnetic waves is soon depressed at high harmonics.
The main effect of the coupling is to further depress the growth of 
electromagnetic waves, whereas the growth-rate of magnetosonic waves
stays about the same.

In the lower panels of Fig.\ref{fig:disprel},  we show the solution
of the dispersion equation in cases when the ion component is not treated as 
perfectly cold at the onset of the instability. The plasma composition is 
the same
as above but now a thermal spread of the ion distribution is included, with
the distribution function in Eq.\ref{eq:ring} replaced by:
\begin{equation}
f_i(u_\perp,u_\parallel)={\delta(u_\parallel) \over 2 \pi}\ 
{\exp[-(u_\perp-u_{0})^2/\Delta u^2] \over \sqrt{\pi}\ \Delta u\ u_0}\ .
\label{eq:warmion}
\end{equation}
The effect of the thermal dispersion of the momenta depends on the effective
Mach number $M_s \equiv u_{0i} / \Delta u $; we use this terminology, anticipating the
occurrence of the ion cyclotron instability in the magnetosonic shock structures studied in 
\S \ref{sec:shocks}, where $u_{0i}$ becomes the upstream 4-velocity and $\Delta u$ the
upstream dispersion in the ion 4-velocity. We assume $M_s=0.01$ in the left panel, while the case 
$M_s=0.1$ is on the right. It is apparent that the growth of the waves 
is suppressed at high harmonic numbers, with the suppression affecting 
increasingly low harmonics the larger the distribution in perpendicular
momentum space.

\end{document}